\documentclass[twocolumn]{aastex631}

\usepackage{url}
\usepackage{amsmath}
\usepackage{booktabs}
\usepackage{graphicx}
\usepackage{xspace}
\usepackage{amsmath}
\let\oldAA\AA
\renewcommand{\AA}{{\text{\normalfont\oldAA}}}
\newcommand{\eg}{\textit{e.g.}}
\newcommand{\ie}{\textit{i.e.}}

\shorttitle{Diverse Ly$\alpha$ Nebulae}
\shortauthors{M.Li et al.}
\graphicspath{{./}{figures/}{figures/postage-stamp/}}
\newcommand{\titleAlpha}{\texorpdfstring{$\alpha$\xspace}{α\xspace}}

\begin{document}

\title{MAMMOTH-Subaru. II. Diverse Populations of Circumgalactic Ly$\alpha$ Nebulae at Cosmic Noon}

\suppressAffiliations 

\correspondingauthor{}
\email{zcai@mail.tsinghua.edu.cn}
\email{lmytime@hotmail.com}
\email{hbz@icrr.u-tokyo.ac.jp}

\author[0000-0001-6251-649X]{Mingyu Li}
\affiliation{Department of Astronomy, Tsinghua University, Beijing 100084, People’s Republic of China}

\author[0000-0003-2273-9415]{Haibin Zhang}
\affiliation{Department of Astronomy, Tsinghua University, Beijing 100084, People’s Republic of China}
\affiliation{National Astronomical Observatory of Japan, 2-21-1 Osawa, Mitaka, Tokyo 181-8588, Japan}

\author[0000-0001-8467-6478]{Zheng Cai}
\affiliation{Department of Astronomy, Tsinghua University, Beijing 100084, People’s Republic of China}

\author[0000-0002-2725-302X]{Yongming Liang}
\affiliation{Institute for Cosmic Ray Research, The University of Tokyo, Kashiwa, Chiba 277-8582, Japan}

\author[0000-0003-3954-4219]{Nobunari Kashikawa}
\affiliation{National Astronomical Observatory of Japan, 2-21-1 Osawa, Mitaka, Tokyo 181-8588, Japan}
\affiliation{Department of Astronomy, School of Science, The University of Tokyo, 7-3-1 Hongo, Bunkyo-ku, Tokyo 113-0033, Japan}
\affiliation{Research Center for the Early Universe, The University of Tokyo, 7-3-1 Hongo, Bunkyo-ku, Tokyo 113-0033, Japan}

\author[0000-0002-0564-891X]{Ke Ma}
\affiliation{Department of Astronomy, Tsinghua University, Beijing 100084, People’s Republic of China}

\author[0000-0003-3310-0131]{Xiaohui Fan}
\affiliation{Steward Observatory, University of Arizona, 933 North Cherry Avenue, Tucson, AZ 85721, USA}

\author[0000-0002-7738-6875]{J. Xavier Prochaska}
\affiliation{Department of Astronomy \& Astrophysics, UCO/Lick Observatory, University of California, 1156 High Street, Santa Cruz, CA 95064, USA}
\affiliation{Kavli Institute for the Physics and Mathematics of the Universe (Kavli IPMU), 5-1-5 Kashiwanoha, Kashiwa, 277-8583, Japan}

\author[0000-0003-2983-815X]{Bjorn H. C. Emonts}
\affiliation{National Radio Astronomy Observatory, 520 Edgemont Road, Charlottesville, VA 22903, USA}

\author[0000-0002-9373-3865]{Xin Wang}
\affiliation{School of Astronomy and Space Science, University of Chinese Academy of Sciences (UCAS), Beijing 100049, China}
\affiliation{National Astronomical Observatories, Chinese Academy of Sciences, Beijing 100101, China}
\affiliation{Institute for Frontiers in Astronomy and Astrophysics, Beijing Normal University, Beijing 102206, China}

\author[0000-0003-0111-8249]{Yunjing Wu}
\affiliation{Department of Astronomy, Tsinghua University, Beijing 100084, People’s Republic of China}
\affiliation{Steward Observatory, University of Arizona, 933 North Cherry Avenue, Tucson, AZ 85721, USA}

\author[0000-0002-0427-9577]{Shiwu Zhang}
\affiliation{Department of Astronomy, Tsinghua University, Beijing 100084, People’s Republic of China}
\affiliation{Research Center for Astronomical Computing, Zhejiang Laboratory, Hangzhou 311100, China}

\author[0000-0002-3119-9003]{Qiong Li}
\affiliation{Jodrell Bank Centre for Astrophysics, University of Manchester, Oxford Road, Manchester, UK}

\author[0000-0001-9487-8583]{Sean D. Johnson}
\affiliation{Department of Astronomy, University of Michigan, Ann Arbor, MI 48109, USA}

\author[0000-0002-5367-8021]{Minghao Yue}
\affiliation{MIT Kavli Institute for Astrophysics and Space Research, 77 Massachusetts Ave., Cambridge, MA 02139, USA}
\affiliation{Steward Observatory, University of Arizona, 933 North Cherry Avenue, Tucson, AZ 85721, USA}

\author[0000-0002-4770-6137]{Fabrizio Arrigoni Battaia}
\affiliation{Max-Planck-Institut f\"ur Astrophysik, Karl-Schwarzschild-Straße 1, D-85748 Garching bei M\"unchen, Germany}

\author[0000-0001-5804-1428]{Sebastiano Cantalupo}
\affiliation{Department of Physics, University of Milan Bicocca, Piazza della Scienza 3, 20126, Milano, Italy}

\author[0000-0002-7054-4332]{Joseph F. Hennawi}
\affiliation{Leiden Observatory, Leiden University, P.O. Box 9513, 2300 RA Leiden, The Netherlands}
\affiliation{Department of Physics, Broida Hall, UC Santa Barbara, Santa Barbara, CA 93106-9530, USA}

\author[0000-0003-3214-9128]{Satoshi Kikuta}
\affiliation{Department of Astronomy, School of Science, The University of Tokyo, 7-3-1 Hongo, Bunkyo-ku, Tokyo 113-0033, Japan}

\author[0000-0001-9442-1217]{Yuanhang Ning}
\affiliation{Department of Astronomy, Tsinghua University, Beijing 100084, People’s Republic of China}

\author[0000-0002-1049-6658]{Masami Ouchi}
\affiliation{Institute for Cosmic Ray Research, The University of Tokyo, Kashiwa, Chiba 277-8582, Japan}
\affiliation{National Astronomical Observatory of Japan, 2-21-1 Osawa, Mitaka, Tokyo 181-8588, Japan}
\affiliation{Kavli Institute for the Physics and Mathematics of the Universe (Kavli IPMU, WPI), University of Tokyo, Kashiwa, Chiba 277-8583, Japan}

\author[0000-0003-4442-2750]{Rhythm Shimakawa}
\affiliation{Waseda Institute for Advanced Study (WIAS), Waseda University, 1-21-1 Nishi-Waseda, Shinjuku, Tokyo 169-0051, Japan}

\author[0000-0003-4877-1659]{Ben Wang}
\affiliation{Department of Astronomy, Tsinghua University, Beijing 100084, People’s Republic of China}
\affiliation{Leiden Observatory, Leiden University, P.O. Box 9513, 2300 RA Leiden, The Netherlands}

\author[0000-0002-9593-8274]{Weichen Wang}
\affiliation{Department of Physics, University of Milan Bicocca, Piazza della Scienza 3, 20126, Milano, Italy}

\author[0000-0003-1887-6732]{Zheng Zheng}
\affiliation{Department of Physics and Astronomy, University of Utah, 115 South 1400 East, Salt Lake City, UT 84112, USA}

\author[0000-0002-9634-2923]{Zhen-Ya Zheng}
\affiliation{Key Laboratory for Research in Galaxies and Cosmology, Shanghai Astronomical Observatory, Chinese Academy of Sciences, 80 Nandan Road, Shanghai 200030, People’s Republic of China}

\begin{abstract}
Circumgalactic Lyman-alpha (Ly$\alpha$) nebulae are gaseous halos around galaxies exhibiting luminous extended Ly$\alpha$ emission.
This work investigates Ly$\alpha$ nebulae from deep imaging of $\sim12~\mathrm{deg}^2$ sky, targeted by the MAMMOTH-Subaru survey.
Utilizing the wide-field capability of Hyper Suprime-Cam (HSC), we present one of the largest blind Ly$\alpha$ nebula selections, including QSO nebulae, Ly$\alpha$ blobs, and radio galaxy nebulae down to typical $2\sigma$ Ly$\alpha$ surface brightness of $(5-10)\times10^{-18}\mathrm{~erg~s^{-1}~cm^{-2}~arcsec^{-2}}$.
The sample contains 117 nebulae with Ly$\alpha$ sizes of 40 - 400 kpc, and the most gigantic one spans about 365 kpc, referred to as the Ivory Nebula.
Combining multiwavelength data, we investigate diverse nebula populations and associated galaxies.
We find a small fraction of Ly$\alpha$ nebulae have QSOs ($\sim7\%$), luminous infrared galaxies ($\sim1\%$), and radio galaxies ($\sim 2\%$).
Remarkably, among the 28 enormous Ly$\alpha$ nebulae (ELANe) exceeding 100 kpc, about 80\% are associated with UV-faint galaxies ($M_\mathrm{UV} > -22$), categorized as Type II ELANe.
We underscore that Type II ELANe constitute the majority but remain largely hidden in current galaxy and QSO surveys.
Dusty starburst and obscured AGN activity are proposed to explain the nature of Type II ELANe.
The SED of stacking all Ly$\alpha$ nebulae also reveals signs of massive dusty star-forming galaxies with obscured AGNs.
We propose a model to explain the dusty nature where the diverse populations of Ly$\alpha$ nebulae capture massive galaxies at different evolutionary stages undergoing violent assembling.
Ly$\alpha$ nebulae provide critical insights into the formation and evolution of today's massive cluster galaxies at cosmic noon.

\end{abstract}

\section{Introduction} \label{sec:intro}

Hydrogen Lyman-alpha (Ly$\alpha$) nebulae are gaseous halos of extended Ly$\alpha$ emission around galaxies, tracing enormous reservoirs of hydrogen gas out to the circumgalactic medium \citep[CGM,][]{Tumlinson2017, Faucher-Giguere2023} and intergalactic medium (IGM) scales.
Ly$\alpha$ nebulae were first detected as giant halos around high-redshift radio galaxies \citep[HzRGs;][]{McCarthy1987} and quasi-stellar objects \citep[QSOs;][]{Hu1987, Heckman1991b}.
The subsequent discovery referred to as Ly$\alpha$ blobs \citep{Steidel2000}, lacking bright ultraviolet (UV) continuum or radio sources, represented a distinct category.
The mysterious power mechanisms of blobs have attracted significant observational and theoretical work aimed at identifying larger samples \citep[\eg,][]{Matsuda2004, Palunas2004, Saito2006, Yang2009} and explaining the origin of their luminous Ly$\alpha$ emission \citep[\eg,][]{Villar-Martin1997, Taniguchi2000, Cantalupo2005, Fardal2001}.
The study of Ly$\alpha$ nebulae has since expanded to encompass diverse populations. 

With the development of modern observing facilities and instruments, Ly$\alpha$ nebulae have been discovered across cosmic time with diverse observed properties \citep{Schirmer2016, Barger2012, Zhang2020, Farina2019}.
In general, Ly$\alpha$ nebulae are a rare population of sources, having high Ly$\alpha$ luminosities ($L_{\rm Ly\alpha} \gtrsim 10^{43} \mathrm{~erg~s^{-1}}$) and large spatially extended Ly$\alpha$ emitting regions ($\gtrsim 40 \mathrm{~kpc}$).
An extreme subclass called enormous Ly$\alpha$ nebulae (ELANe) has been revealed at cosmic noon ($z\simeq 2-3$) in past decades \citep[\eg,][]{vanOjik1997, Reuland2003, Villar-Martin2003, Miley2006, Cantalupo2014, Hennawi2015, Cai2017a, ArrigoniBattaia2018a}.
ELANe represent the extremes of Ly$\alpha$ nebulosities with higher Ly$\alpha$ luminosity ($L_{Ly\alpha} \gtrsim 10^{44} \mathrm{~erg~s^{-1}}$) and larger extent ($\gtrsim 100 \mathrm{~kpc}$).
The detection of such enormous Ly$\alpha$ nebulae provides unique insights into massive CGM gas flows and accretion at the peak epoch of galaxy growth. 

A few studies use Ly$\alpha$ nebulae to examine the galactic ``ecosystem'' comprising IGM, CGM, embedded galaxies, and supermassive black holes (SMBHs).
They provide direct observational access to the densest part of the IGM and CGM, enabling studies of gas accretion, feedback, and recycling \citep[\eg,][]{ArrigoniBattaia2018a, Ao2020, Wang2021, Li2021b, Daddi2021, Zhang2023c, Zhang2024}.
In contrast to these Ly$\alpha$ nebulae, however, in most other systems, the CGM/IGM is usually too faint to be visible in emission and can only be detected through absorption along just a few sightlines \citep[\eg,][]{Zhu2013, Napolitano2023, Zou2024, Wu2023b}.
Ly$\alpha$ nebulae, therefore, represent rare and unique opportunities to survey the formation and evolution of the galaxy in the early universe \citep{Ouchi2020}.

Several observational techniques have been utilized to identify Ly$\alpha$ nebulae.
Narrowband imaging surveys are the first and remain one of the most effective approaches \citep[\eg,][]{Fynbo1999, Keel1999, Francis2001, Palunas2004, Dey2005, Nilsson2006, Smith2007b, Ouchi2008, Matsuda2009, Yang2010, Erb2011, Matsuda2011, Cantalupo2012, Kikuta2017, Shibuya2018b, Kikuta2019, Rahna2022, Ramakrishnan2023}.
Similarly, medium-band and broad-band imaging have also proved useful for finding luminous Ly$\alpha$ nebulae based on color excess \citep[\eg,][]{Saito2006, Prescott2012a, Barger2012, Hibon2020, Shimakawa2022}.
More recently, integral field spectrograph/unit (IFS/IFU) instruments on 8-10m class telescopes, such as VLT/MUSE and Keck/KCWI, have enabled the most comprehensive studies of Ly$\alpha$ nebulae both for blind but small field-of-view surveys \citep[\eg,][]{Caminha2016, Vanzella2016, Leclercq2017, Kusakabe2022} and targeted follow-up of pre-selected sources \citep[\eg,][]{Borisova2016, ArrigoniBattaia2019, Cai2019}.

Nevertheless, most of the discovered nebula samples still rely on pre-selections of individual targets, which may lead to an incomplete picture of the whole Ly$\alpha$ nebulae population.
Ly$\alpha$ nebulae are thought to relate to the formation of massive galaxies whose dark matter halo has a typical mass of $\sim 10^{13}{\rm M_\odot}$, living in the most active star formation region and trace large-scale mass overdensities.
Therefore, one successful pre-selection technique is to target biased tracers of massive halos, such as luminous QSOs \citep[\eg,][]{Borisova2016, ArrigoniBattaia2016, ArrigoniBattaia2019, Cai2019, Farina2019, Mackenzie2021}, high-redshift radio galaxies \citep[\eg,][]{Swinbank2015, Wang2021}, active galactic nucleus (AGN) pairs/groups \citep[\eg,][]{Hennawi2015, Cai2018}, galaxy groups/clusters \citep[\eg,][]{Daddi2021, Daddi2022, Jimenez-Andrade2023}, and dust-obscured galaxies \citep[\eg,][]{Bridge2013, Ginolfi2022}.

A wide-field survey independent of individual target pre-selections is needed to study the diverse population of Ly$\alpha$ nebulae comprehensively.
Compared to blind fields, Ly$\alpha$ nebulae, especially ELANe, tend to reside in high-redshift overdense regions \citep[\eg,][]{Steidel2000, Dey2005, Matsuda2005, Prescott2009, Yang2010, Hennawi2015, Cai2017a, Cai2017b, Cai2018, ArrigoniBattaia2018a}.
Therefore, the effects of nebula clustering in overdensity will enable the discovery of more and larger Ly$\alpha$ nebulae.

Based on the selection with QSO spectra as sightlines \citep[\eg,][]{Cai2017a}, we construct a sample of large-scale structures at $z=2-3$, using IGM Ly$\alpha$ forest absorption, \ie, coherently strong Ly$\alpha$ absorption system (CoSLA) as tracers \citep{Cai2016}. 
Pilot surveys successfully identify a few massive protoclusters at $z=2-3$, \eg, BOSS1441 \citep{Cai2017b}, BOSS1244, and BOSS1542 \citep{Zheng2020, Shi2021}.
An ELAN dubbed MAMMOTH-1 \citep{Cai2017a} was discovered close to the density peak of protocluster BOSS1441, which has a spatial extent of over 400 kpc and the highest extended Ly$\alpha$ luminosity to date.
Such an extreme ELAN provides a great opportunity to systematically study CGM gas flows directly to reveal aspects of galaxy evolution.

Encouraged by the successful pilot study, we have constructed a larger sample of eight MAMMOTH fields using Subaru Hyper Suprime-Cam \citep[Subaru/HSC;][]{Miyazaki2018}, and this project is named MAMMOTH-Subaru (also see \citealt[][Liang et al. in prep.]{Liang2021, Zhang2023a, Zhang2024, Ma2024, Liang2024}).
Thanks to the sufficiently wide field of view (FoV) of HSC (1.5$^\circ$ diameter), the deep narrowband imaging in these fields allows us to construct a complete sample of Ly$\alpha$ nebulae, without the need for pre-selection of powering sources (e.g., Type-I QSO). 
Therefore, our sample can represent a comprehensive view of the Ly$\alpha$ nebula population powered by various sources and mechanisms. 

In this paper, we investigate a large sample of Ly$\alpha$ nebulae at $z\sim2.2 - 2.3$, selected from a wide, deep narrowband imaging survey.
The structure of this paper is as follows:
In $\S$\ref{sec:obs}, we describe our observations and data reduction procedures.
In $\S$\ref{sec:selection}, we present the sample selection for Ly$\alpha$ nebulae.
In $\S$\ref{sec:res}, we report the nebula sample, show the results of observed diversity in multi-wavelength properties, and propose a picture to explain the diverse population.
In $\S$\ref{sec:summary}, we summarize this work and present brief conclusions.
In this work, we use the vacuum wavelength $1215.670$\AA\ for Ly$\alpha$ based on the Atomic Line List v2.04\footnote{\url{https://linelist.pa.uky.edu/atomic/index.html}}.
Throughout this paper, magnitudes are given in the AB system \citep[\eg,][]{Oke1974, Oke1983} unless otherwise specified.
We adopt a flat $\mathrm{\Lambda CDM}$ cosmology with $\Omega_{\mathrm{m},0}=0.3$ and $H_{0}=70 \mathrm{~km}\mathrm{~s}^{-1} \mathrm{~Mpc}^{-1}$.
In this cosmology, at the redshift $z=2.18$, 1\arcsec\ corresponds to a physical 8.28 kpc; at the redshift $z=2.29$, 1\arcsec\ corresponds to 8.21 physical kpc.

\begin{figure}
\centering
\includegraphics[width=\linewidth]{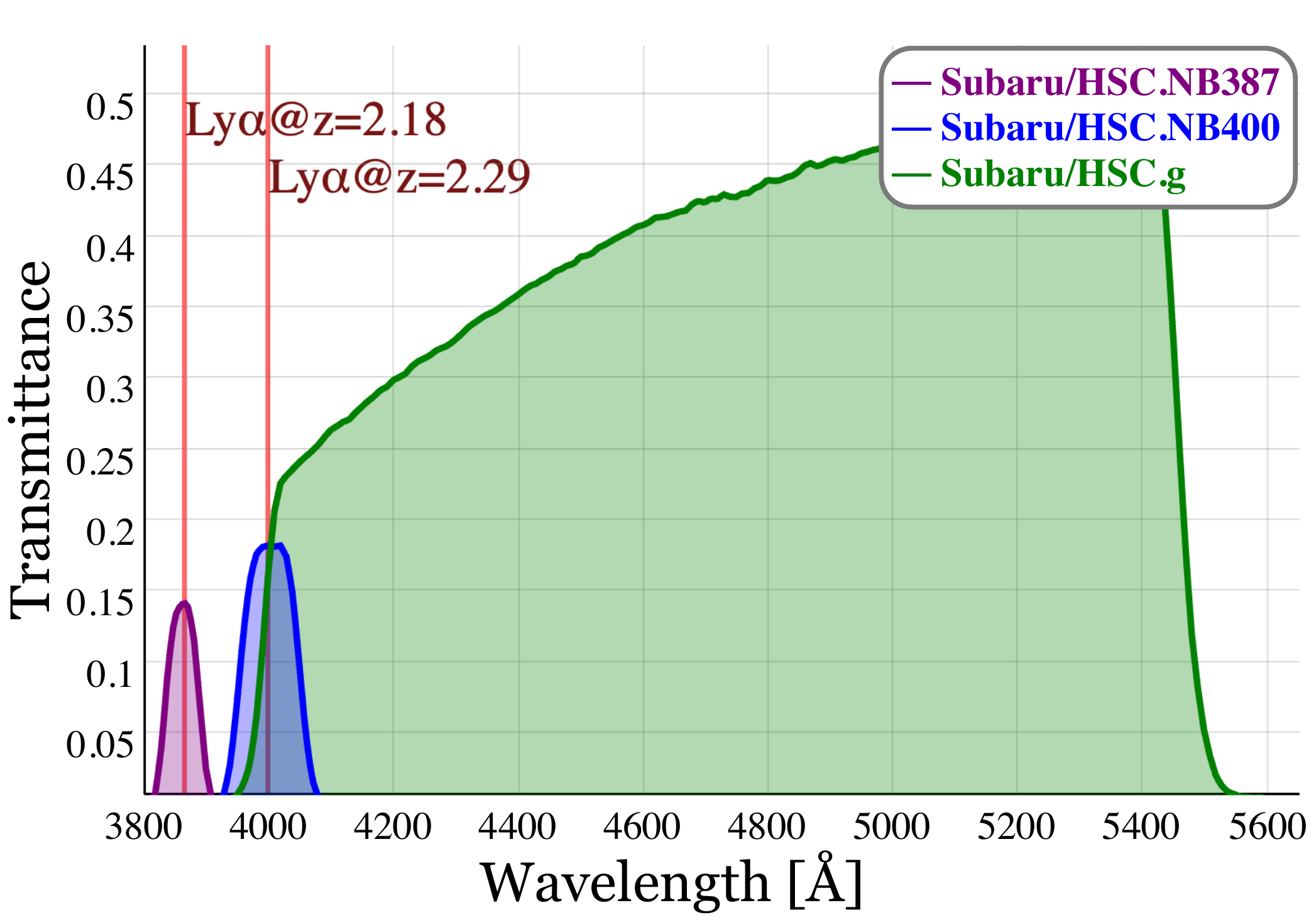}
\caption{Transmission curves of NB387, NB400, and g band mounted on the HSC. The purple, blue, and green solid lines (from left to right) represent the total transmittance of NB387, NB400, and g band respectively. Two red vertical lines denote the wavelength of the redshifted Ly$\alpha$ line at $z=2.18$ and $z=2.29$. This figure is generated by our interactive web application \href{https://github.com/lmytime/MyFilter}{\texttt{MyFilter}}.
\label{fig:filter}}
\end{figure}

\section{Observations and Data} \label{sec:obs}
Our survey covers eight SDSS/(e)BOSS fields (see Table~\ref{tab:field}) selected based on the MAMMOTH approach and/or QSO clustering (\citealt[][Liang et al. in prep.]{Cai2016, Liang2021, Liang2024}).
The Subaru observations took deep narrowband and broadband imaging for these eight fields.
Furthermore, we performed spectroscopic observations for a few selected Ly$\alpha$ nebulae. 
The observation details are as follows.

\begin{deluxetable*}{cccccccccc}[ht]
\tablenum{1}
\tablecaption{Summary of field information\label{tab:field}}
\tablewidth{0pt}
\tablehead{
\colhead{Field} & \colhead{R.A.} & \colhead{Decl.} & \colhead{Obs Period} & \colhead{$\theta_\mathrm{NB}$} & \colhead{$\theta_\mathrm{BB}$} & \colhead{$m_{\mathrm{NB}, 5 \sigma}$} & \colhead{$m_{\mathrm{BB}, 5 \sigma}$} & \colhead{Area} & \colhead{Ly$\alpha$ SB$_{2\sigma}$} \\
\colhead{name} & \colhead{hh:mm:ss.ss} & \colhead{dd:mm:ss.s} & \colhead{month, year} &
\colhead{arcsec} & \colhead{arcsec} & \colhead{mag} & \colhead{mag} & \colhead{deg$^2$}& \colhead{$10^{-18}$~cgs.$~\rm arcsec^{-2}$}\\
\colhead{(1)} & \colhead{(2)} & \colhead{(3)} & \colhead{(4)} &
\colhead{(5)} & \colhead{(6)} & \colhead{(7)} & \colhead{(8)} & \colhead{(9)} & \colhead{(10)}
}
\startdata
J0210 & 02:09:58.90 & +00:53:43.0 & Jan., 2018      & 1.22   & 0.90   & 24.25     & 26.34  & 1.34 & 10.56  \\
J0222 & 02:22:24.66 & -02:23:41.2 & Jan., 2018      & 0.90   & 0.90   & 24.99     & 27.01  & 1.13 & 8.23   \\
J0924 & 09:24:00.70 & +15:04:16.7 & Jan. \& Mar., 2019 & 0.84   & 0.79   & 24.74     & 26.63  & 1.47 & 9.79    \\
J1419 & 14:19:33.80 & +05:00:17.2 & Mar., 2019      & 0.86   & 0.70   & 24.81     & 26.80  & 1.45   & 10.66  \\
J0240 & 02:40:05.11 & -05:21:06.7 & Nov., 2019        & 1.00   & 0.84   & 25.61     & 26.80  & 1.53  & 7.78   \\
J0755 & 07:55:35.89 & +31:09:56.9  & Nov., 2019     & 0.88   & 1.16   & 25.83     & 26.50   & 1.54  & 5.04  \\
J1133 & 11:33:02.40 & +10:05:06.0  &  Mar., 2020    & 0.85   & 0.82   & 25.49     & 26.30   & 1.55 & 9.19   \\
J1349 & 13:49:40.80 & +24:28:48.0  &  Mar., 2020        & 0.98   & 0.86   & 25.67     & 26.15   & 1.62 & 7.76 \\
\enddata
\tablecomments{(1) the name of fields; (2-3) field center coordinates Right Ascension and Declination in equinox with an epoch of J2000; (4) the period of executing the observations; (5-6) the FWHMs of PSFs measured by \texttt{PSFEx}; (7-8) are the $5 \sigma$ limiting magnitudes measured in an aperture with the radius of $1.7^{\prime \prime}$ ($2.5^{\prime \prime}$ for J0210 field to match the seeing) for the final stacked NB and $g$-band image, respectively; (9) the effective survey area after masking; (10) the average $2\sigma$ detection limit of Ly$\alpha$ line map in unit of $10^{-18}\mathrm{erg~s^{-1}~cm^{-2}~arcsec^{-2}}$.}
\end{deluxetable*}

\subsection{Imaging Observations}
In this section, we summarize the Subaru imaging observations in programs S17B-041 (PI: N. Kashikawa), S19A-TE288, S19B-ChinaTAP (PI: Z. Cai), and S20A-108 (PI: Y. Liang), but please see \citet{Liang2021} and Liang et al. (in prep.) for details. 

In our survey, we perform deep narrowband imaging using NB387 ($\rm \lambda_c$=3863\AA, $\Delta \lambda$=55\AA) and NB400 ($ \rm \lambda_c$= 4003\AA, $\Delta \lambda$=92\AA) filters on the HSC, which covers $1.5^\circ$ FoV in diameter.
The narrowband filters enable us to detect Ly$\alpha$ emission at the corresponding redshifts of $z=2.177\pm0.023$ and $z=2.293 \pm 0.038$, respectively.
The line-of-sight depth of the redshift slice is 62.3 cMpc for NB387 and 97.8 cMpc for NB400.
The HSC-g band ($\lambda_c$= 4754\AA, $\Delta \lambda$=1395\AA) is used for the continuum estimate of the line-emitting sources.
The transmission curves of these three filters are shown in Figure~\ref{fig:filter}, which is generated by our interactive web application \texttt{MyFilter}\footnote{\url{https://github.com/lmytime/MyFilter}}\citep{MyFilter}.

The HSC imaging data were reduced using the HSC pipeline, \texttt{hscPipe} \citep{Bosch2018}.
The details of imaging reductions can be checked in \citet{Liang2021}.
Here we present a brief summary of the reduction procedures.
First, the primary calibration data (\ie, the bias, dark, flat, and global sky) were made.
Then they were applied to each visit.
Pan-STARRS DR1 \citep{Flewelling2020} catalogs are regarded as references for astrometry and photometry.
Finally, the coadding process was performed for each visit to generate mosaic image products.
Note that systematic photometry corrections due to the stellar metallicity bias were included in our reduction, which addresses a known issue also reported in the HSC Subaru Strategic Program \citep[SSP;][]{Aihara2019, Aihara2022}.

\subsection{Spectroscopic Observations}\label{sec:spec-obs}
We take follow-up spectroscopic observations for a few selected Ly$\alpha$ nebula candidates (see section \S\ref{sec:selection} for the selection) using the MMT/Binospec spectrograph \citep{Fabricant2019} and the Magellan/IMACS spectrograph \citep{Dressler2011}.

The observations of MMT/Binospec are carried out in April 2021 and July 2022.
We use the long-slit mode of Binospec with 1\arcsec\ width and 2$^\prime$ length to take the spectra.
Different instrument configurations are used for April and July observations.
In the April observations, we adopt the LP3500 filter and the 1000 lines/mm grating (1000~lpm mode hereafter), which allows photons from 3720 {\AA} to 4950 {\AA} with the spectral resolution of $R\simeq 3900$.
Each source is taken for $\rm 10\times1200 \mathrm{~s} = 3.3\mathrm{~hrs}$.
In the July observations, we use the LP3800 filter and 600 lines/mm grating (600~lpm mode hereafter), covering from 4000 {\AA} to 6400 {\AA} with $R\simeq2740$.
The integration time is $\rm 4\times1200 \mathrm{~s} = 1.3\mathrm{~hrs}$.
The Binospec setting allows coverage of the redshifted emission lines, including Ly$\alpha$ and \ion{N}{5} for sources at $z\simeq2$.
We reduce the Binospec data using the official \texttt{IDL} pipeline \citep{Kansky2019b}.

The observations of Magellan/IMACS are taken on September 29 and 30, 2022, using the multi-slit spectroscopy mode with the f/2 camera. The field of view of IMACS is a circle with a radius of $12\arcsec$.
We use the 400 lines per mm grism combined with a custom filter to cover the wavelength between $\simeq 3600$ {\AA} and 5700~\AA.
The spectral resolution is $\sim7$~{\AA} using this setup with a slit width of 1.2 arcsec and slit length of 8.0 arcsec.
We observe one pointing in each of the J0210 and J0222 fields for NB387 sources, and two pointings in the J0240 field for NB400 sources.
The on-source exposure times are 7500 s for NB387 fields and 6000 s for NB400 fields.
We reduce the spectroscopic data using the official pipeline \texttt{COSMOS} \citep{Oemler2017}.

\section{Sample Selection} \label{sec:selection}
The Ly$\alpha$ nebula samples are selected through three steps:
(1) constructing source detection and photometry catalog;
(2) selecting emitters based on color excess of the narrow-band (NB) compared to the broad-band (BB);
(3) identifying nebulae based on their extended area.
While similar to the Ly$\alpha$ emitter (LAE) selection with the same HSC images \citep{Liang2021, Zhang2024}, the procedure for searching for Ly$\alpha$ nebulae requires some optimal adjustments of source detection, which are detailed in this section. 

\subsection{Photometry Catalog and Emitter Selection}
We construct photometric catalogs in the narrow-band (NB387 or NB400) and broad-band (g-band) mainly using \texttt{SExtractor 2.25.0} \citep{Bertin1996}.
Firstly, we use \texttt{Astropy} to crop and correct the NB and BB images to the same shape and refine them to the same world coordinate system (WCS).
We run \texttt{SExtractor} in dual-image mode with the narrow-band images as detection references.
The detection threshold is set as 15 continuous pixels over the $2\sigma$ sky background.
To match the seeing, we adopt a $7\times7$ convolution mask of a Gaussian kernel with FWHM = 3.0 pixels to smooth the image before source detection.
We apply the sky background root-mean-square (RMS) map as the weighting map to minimize the slight image depth fluctuation effect.
Besides, we set the global background mode with a mesh size of 128 pixels.
Finally, the narrow-band and broad-band source catalogs are combined naturally.
We manually make masks to remove the sources in the regions around saturated stars, with severe artifact contamination and with severe stray lights.
Please check Table~\ref{tab:field} for the field survey areas of our eight fields after masking.
When the objects in the broadband are fainter than the $2\sigma$ limiting magnitude (see Table~\ref{tab:field}), we replace their magnitude with the corresponding $2\sigma$ limiting magnitude for color excess analysis.

We simultaneously adopt Kron-like elliptical aperture magnitudes (\texttt{MAG\_AUTO}) and circular aperture magnitudes (\texttt{MAG\_APER}) to derive photometric properties.
The aperture radii are set to be $10~\mathrm{pixels}\simeq1.7\arcsec$ ($15~\mathrm{pixels}\simeq2.5\arcsec$ for the J0210 field due to the poor seeing) to match the seeing.
The point spread function (PSF) is measured when constructing the photometry catalog using \texttt{PSFEx 3.17.1} \citep{Bertin2011}.
The FWHM value of the PSF is used for the PSF-matching procedure between the narrowband and broadband images.

Based on the detection and photometry catalog constructed using the methods shown above, we apply the same selection criteria as shown in \citealt{Liang2021, Zhang2024} to select emitters with a rest-frame Ly$\alpha$ equivalent width $\rm EW_0 > 20$\AA.
It should be noted that the criteria may select a few unmasked stray lights, hot pixels, moving objects, saturated objects, and other artifacts.
Visual inspection is conducted to eliminate them.
For see further details on emitter selection, please refer to \citet{Ma2024} and \citet{Zhang2024}.

\begin{figure*}
\centering
\includegraphics[width=\linewidth]{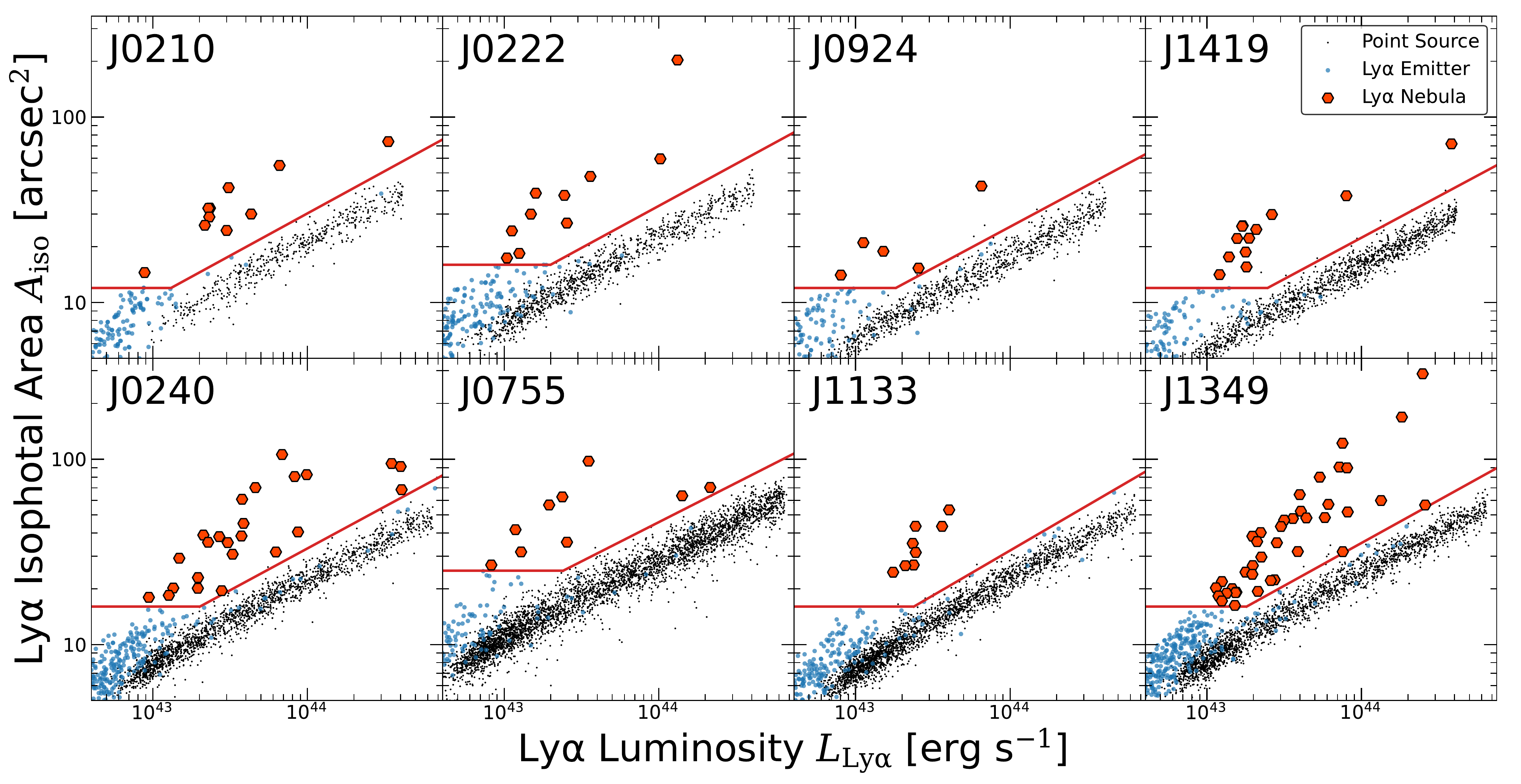}
\caption{Isophotal areas ($A_\mathrm{iso}$) versus Ly$\alpha$ luminosity ($L_\mathrm{Ly\alpha}$) of selected Ly$\alpha$ nebulae in our eight fields. Blue dots represent Ly$\alpha$ emitters (LAEs), and red hexagons denote selected nebulae. The inclined lines denote the $3\sigma$ above the sequence of point sources (black dots), indicating that the sources above have spatially extended profiles. The nebula selection criteria are bounded to the upper side by the red lines. Our selected nebulae are located well above this relation.
\label{fig:area-lum}}
\end{figure*}

\begin{figure*}[ht]
\centering
\includegraphics[width=\linewidth]{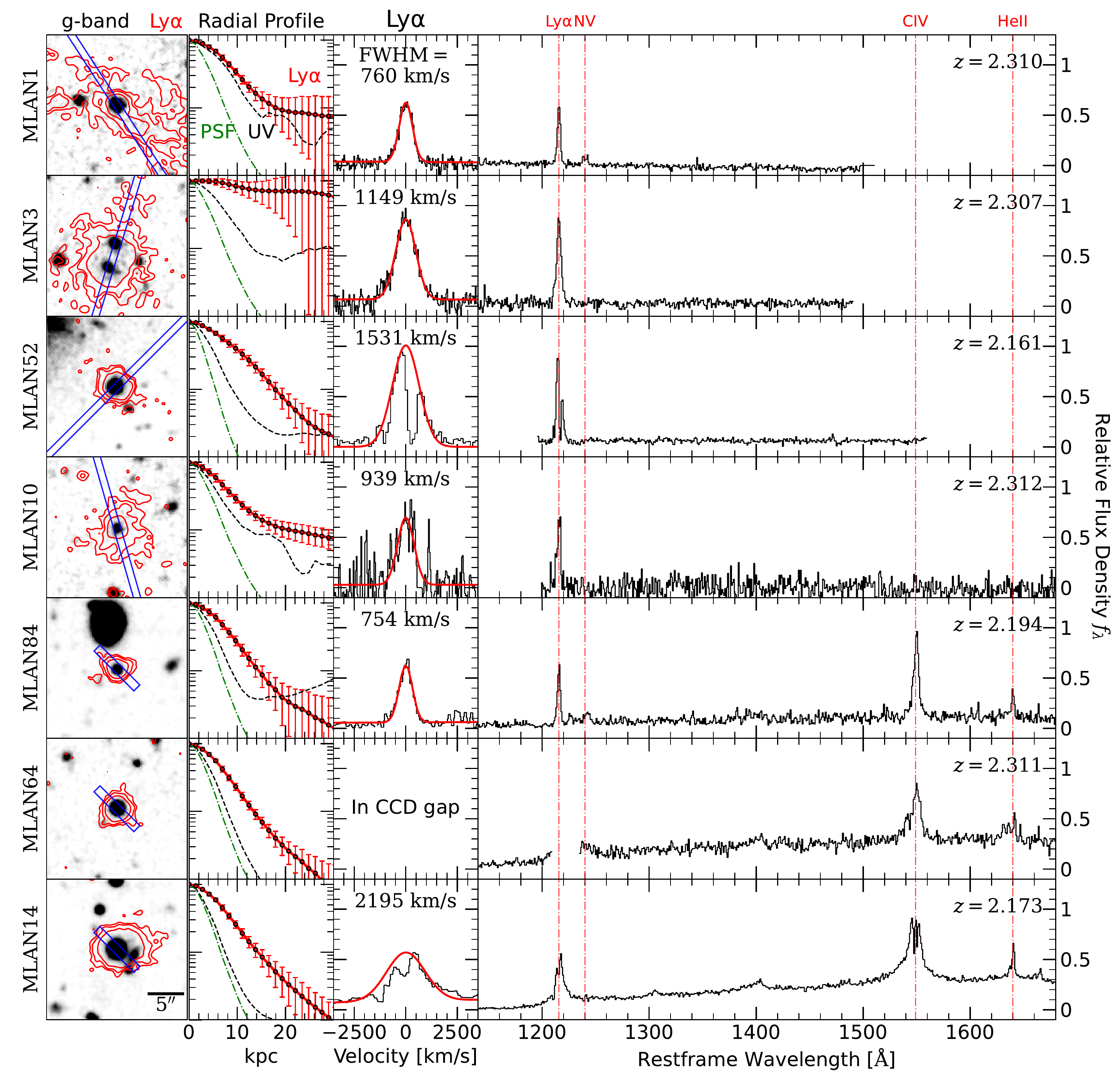}
\caption{Cutout images, radial profile, and spectra of spectroscopically confirmed Ly$\alpha$ nebulae. Each row presents one object. For each object, the left panel shows the g-band image with overlaid $3,5,10\sigma$ Ly$\alpha$ contours (red lines). The blue rectangles denote the positions of spectroscopic slits. The middle-left panels show the surface brightness radial profiles of Ly$\alpha$, UV continua, and PSF with the image center in the left panels as the profile center. The right panels present the extracted 1D spectrum in the rest frame with redshift noted. The middle-right panels zoom into the Ly$\alpha$ wavelength in relative velocity space and red lines denote the best-fit of the Ly$\alpha$ emission lines with FWHM annotated at the top.
\label{fig:spec}}
\end{figure*}

\subsection{Emission Line Map and Nebula Selection}\label{sec:nebula-selection}

We make line emission images (Ly$\alpha$ images) to measure the luminosity and spatial extent of Ly$\alpha$ emission to select nebulae.
Before making the Ly$\alpha$ image, we match the PSF between the NB and BB image using a Gaussian kernel with FWHM of measured PSF (see Table~\ref{tab:field}).
We use a linear combination of the NB and BB images with different coefficients to produce the continuum and Ly$\alpha$ image.
For NB387 emitters, the equations are:
\begin{equation}
\begin{aligned}
f_{\rm \nu, cont} &= \langle f_{\nu} \rangle_\mathrm{g},\\
F_{\rm Ly\alpha} &= \left(\langle f_{\nu} \rangle_\mathrm{NB387} - \langle f_{\nu} \rangle_\mathrm{g} \right) \cdot \frac{c\cdot55.9~\AA}{(3862.0~\AA)^2};\\
\end{aligned}
\end{equation}
and for NB400 emitters, the equations are:
\begin{equation}
\begin{aligned}
f_{\rm \nu, cont} &= 1.027 \langle f_{\nu} \rangle_\mathrm{g} - 0.027 \langle f_{\nu} \rangle_\mathrm{NB400},\\
F_{\rm Ly\alpha} &= \left(\langle f_{\nu} \rangle_\mathrm{NB400} - \langle f_{\nu} \rangle_\mathrm{g} \right) \cdot \frac{c\cdot101.0~\AA}{(4000.9~\AA)^2}.\\
\end{aligned}
\end{equation}
In these equations, $\langle f_{\nu} \rangle_\mathrm{g}$, $\langle f_{\nu} \rangle_\mathrm{NB387}$, and $\langle f_{\nu} \rangle_\mathrm{NB400}$ denote the measured flux density in each filter and $c$ is the light speed in vacuum.
Please refer to Appendix~\ref{sec:estimate_line} for the detailed definition, methodology, and derivation of these equations.

Note that the above equations are based on two assumptions.
One assumption is that the continuum flux density of Ly$\alpha$ emitting sources is $f_\nu = \rm constant$ (or $f_\lambda \propto \lambda^\beta$ where the slope $\beta=-2$).
The other assumption is that we assume the emission line is described by a delta function with the pivot wavelength of the narrowband as the line wavelength.
Different assumed UV slopes could affect the Ly$\alpha$ luminosity estimate, but the effects are tested to be within about 0.1 dex \citep[\eg,][]{Liang2021}.
If the true line wavelength is offset from the pivot wavelength, we will underestimate the line flux.
Therefore, note that the estimated line fluxes are lower limits due to these systematic issues.

We measure the limiting surface brightness for the Ly$\alpha$ images in the eight fields and presented them in Table~\ref{tab:field}.
The typical $2\sigma$ limiting surface brightness is $\mathrm{SB}_{2\sigma} = 5-10\times10^{-18}\mathrm{~erg~s^{-1}~cm^{-2}~arcsec^{-2}}$ measured by the pixel-to-pixel fluctuation.
Based on the Ly$\alpha$ image, we run \texttt{SExtractor} in association mode to measure the isophotal area and isophotal flux of selected LAEs over the $2\sigma$ limiting surface brightness.
We derive the Ly$\alpha$ luminosity from the line flux assuming the Ly$\alpha$ at redshift $z=2.18$ for NB387 and $z=2.29$ for NB400.
To distinguish between point-like Ly$\alpha$ emitting sources and extended nebulae in our observation depth, we require the isophotal area of candidate nebulae to be larger than that of the point sources.
The point-like sources in each field are identified from the half-flux radii $r_{0.5}$ v.s. magnitude diagram.
The half-flux radii of point sources are almost constant and only depend on the FWHM of PSF, leading to horizontal sequences in the $r_{0.5}$ v.s. magnitude diagram.
We select these point sources within the $1\sigma$ confidence level of the sequence.

We use these point-like sources as references to determine whether our selected emitters have extended emission profiles.
Assuming that all the flux in the narrowband for the point source comes from Ly$\alpha$ emission, we can mark these point sources on the area v.s. luminosity figure, and they are arranged in a linear sequence (black dots in Figure~\ref{fig:area-lum}).
The areas of nebulae are required to be $3\sigma$ above this point source sequence, where $\sigma$ is the root mean square error of the linear regression for the sequences.
Note that it is difficult to distinguish the point source and the extended source if the source has faint diffuse emission.
Consequently, we require additional minimal area criteria for a robust nebula selection.
The minimal areas are chosen to match the different detection limits of our eight fields.
The previous literature \citep[\eg,][]{Kimock2021, Yang2009, Yang2010, Badescu2017} gave a few different minimal area criteria despite considering the disparity of depths.
We refer to the simulation of \citet{Kimock2021} to give an estimate of Ly$\alpha$ extended area for nebulae at $z=2$.
The criteria of minimal area are chosen to match the imaging depth as $12\mathrm{~arcsec^2}$ for J0210, J0924, and J1419 fields; $16\mathrm{~arcsec^2}$ for J0222, J0240, J1133, and J1349 fields; $25\mathrm{~arcsec^2}$ for J0755 field.
We adopt tests to fine-tune the minimal area criteria from $10-30\mathrm{~arcsec^2}$, which suggested that our final criteria choice minimized the spurious detection and achieved almost the highest completeness.
The criteria and procedure of nebula selections are shown in Figure~\ref{fig:area-lum}.

We visually inspect these nebulae and checked individual exposures to remove any fake selections.
Fake selections include stray light, moving objects, artifacts, and objects that have been confirmed at $z<2$ \citep[\eg,][]{Flesch2021}.
Note that we exclude five sources that are contaminated by foreground galaxies or stars, whose isophotal areas can be unexpectedly boosted.
Based on our selection and visual inspection, 117 Ly$\alpha$ nebulae are selected in eight fields.
We index the sample according to the Ly$\alpha$ isophotal area (from largest to smallest) in the format of ``MLAN\texttt{id}", where MLAN represents the abbreviation of MAMMOTH Ly$\alpha$ nebula.
The ids, equatorial coordinates, photometry magnitudes, Ly$\alpha$ luminosities, isophotal areas, and end-to-end extents of our nebula sample are shown in Table~\ref{tab:labs}.

\subsection{Spectroscopic confirmation}

We have carried out spectroscopic observations with MMT/Binospec and Magellan/IMACS for seven of our nebula candidates (MLAN1, MLAN3, MLAN10, MLAN52, MLAN64, and MLAN84).
The spectra of four nebulae are taken by Binospec with MLAN1, MLAN3, and MLAN52 in 1000~lpm mode and MLAN10 in 600~lpm mode (see Section~\ref{sec:spec-obs}).
Three nebulae (MLAN14, MLAN84, and MLAN64) are assigned in the slit of the IMACS mask.

Overall, we capture spectra for a total of seven sources.
All these sources are confirmed successfully, and some of them have been detected with other emission lines (\eg, \ion{N}{5}, \ion{C}{4}, and \ion{He}{2}) in addition to Ly$\alpha$.
We use a Gaussian profile to fit the emission lines to determine the redshift based on the best-fit Gaussian centroid.
The redshifts of MLAN14, MLAN64, and MLAN84 are 
based on the \ion{He}{2} line, and the redshifts 
of others are based on the Ly$\alpha$ lines.
For MLAN52, we mask the ``absorption" in the Ly$\alpha$ line to 
fit a Gaussian profile.
The Ly$\alpha$ lines of our confirmed Ly$\alpha$ nebulae have typical FWHM $\gtrsim700~\mathrm{km}$ s$^{-1}$, much larger than the typical LAE Ly$\alpha$ line FWHM $\lesssim400~\mathrm{km}$ s$^{-1}$ \citep[e.g.,][]{Kerutt2022, Ning2020, Ning2022}.
The broad line features, large equivalent width, and coexisting lines indicate that the emission line captured by NB is Ly$\alpha$, and these objects are Ly$\alpha$ nebulae.
The extracted spectra and the best-fit redshifts of these seven sources are shown in Figure~\ref{fig:spec}.
Meanwhile, we also show the cutout images with overlaid Ly$\alpha$ contours and surface brightness radial profiles in Figure~\ref{fig:spec}. 

Note that there may be contamination of other extended line emitters, \eg, [\ion{O}{2}] $\lambda3728$ at $z\simeq0.035$ for NB387 ($z\simeq0.07$ for NB400), \ion{Mg}{2} $\lambda2800$ at $z\simeq0.4$, \ion{C}{3}] $\lambda1908$ at $z\simeq1$, and \ion{C}{4} $\lambda1549$ at $z\simeq1.5$.
However, previous works using a similar selection approach show that the contamination rate is minimal ($\lesssim1\%$) \citep[\eg,][]{Matsuda2004, Yang2010,Badescu2017} because of small survey volume at low-$z$ and the rarity of high EW emission lines \footnote{For example, [\ion{O}{2}] $\lambda3728$ emitters at low-$z$ rarely have rest-frame equivalent width larger than $100\AA$ \cite[\eg,][]{Hogg1998}}.
In addition, the one hundred percent confirmation rate of our spectroscopic observations indicates that our criteria can guarantee a high purity of the sample.

\begin{figure*}[ht]
\centering
\includegraphics[width=\linewidth]{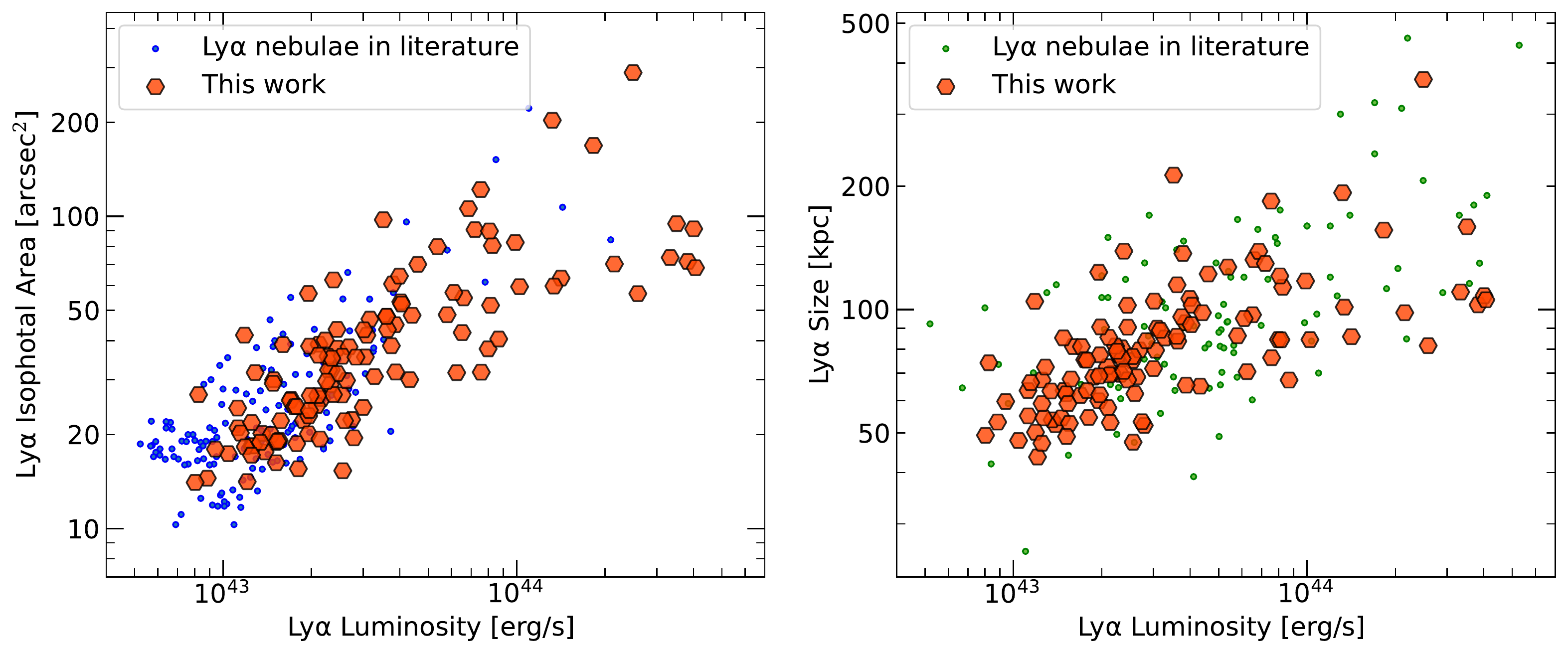}
\caption{Ly$\alpha$ isophotal areas (left) and maximum projected sizes (right) versus Ly$\alpha$ luminosities of the extended Ly$\alpha$ nebulae in this work compared to other samples from the literature. The comparison samples include LABs, QSO nebulae, and ELANe at $z\simeq2-3$ \citep[\eg,][]{Kikuta2019, Yang2010, ArrigoniBattaia2019, Matsuda2004,Matsuda2011, Borisova2016, Badescu2017, Hennawi2015, Cantalupo2014, Cai2017b}. Please check Fig.~\ref{fig:uv-faint} for further classification of diverse nebula populations.
\label{fig:labsum}}
\end{figure*}


\section{Results}\label{sec:res}

\begin{deluxetable*}{ccccccccccc}
\tablenum{2}
\tablecaption{Properties of Ly$\alpha$ Nebulae\label{tab:labs}}
\tablewidth{0pt}
\tablehead{
\colhead{ID} & \colhead{Field} & \colhead{R.A.} & \colhead{Decl.} & \colhead{$m_{\mathrm{NB}}$} & \colhead{$m_{\mathrm{g}}$} & 
 \colhead{$A_{\mathrm{iso}}$} & \colhead{$\log\left(L_{\mathrm{Ly\alpha}}\right)$} & \colhead{Ly$\alpha$ size} & \colhead{$z_\mathrm{spec}$} & \colhead{Comment} \\
\colhead{} & \colhead{} & \colhead{degree} & \colhead{degree} & \colhead{mag} & \colhead{mag} & \colhead{arcsec$^2$} & \colhead{$\mathrm{\log[erg~s^{-1}}]$} & \colhead{kpc} & \colhead{} & \colhead{}
}
\decimalcolnumbers
\startdata
MLAN1 & J1349 & 206.97629 & 23.96409 & 22.68 & 24.08 & 288.7 & 44.40 & 364.7 & 2.310 & T2;\textit{Ivory} \\
MLAN2 & J0222 & 35.18785 & -2.22342 & 23.92 & 25.91 & 203.2 & 44.12 & 192.9 &  & T2 \\
MLAN3 & J1349 & 207.15093 & 23.9955 & 22.89 & 24.55 & 168.7 & 44.26 & 156.3 & 2.307 & T2;R \\
MLAN4 & J1349 & 206.86169 & 24.53714 & 23.43 & 23.35 & 121.9 & 43.88 & 184.0 &  & T2 \\
MLAN5 & J0240 & 39.93041 & -5.99873 & 24.64 & 25.58 & 105.9 & 43.84 & 138.8 &  & T2 \\
MLAN6 & J0755 & 119.53963 & 30.80769 & 25.47 & 26.11 & 97.5 & 43.55 & 212.9 &  & T2 \\
MLAN7 & J0240 & 40.49514 & -5.70917 & 19.85 & 20.35 & 94.7 & 44.54 & 159.2 & 2.267 & Q \\
MLAN8 & J0240 & 39.95441 & -5.32985 & 19.75 & 20.35 & 91.2 & 44.60 & 108.1 & 2.299 & Q \\
MLAN9 & J1349 & 206.98806 & 24.90205 & 23.81 & 24.67 & 90.6 & 43.86 & 129.5 &  & T2 \\
MLAN10 & J1349 & 206.98091 & 23.9814 & 23.00 & 24.77 & 89.8 & 43.91 & 120.8 & 2.312 & T2\\
\multicolumn{11}{c}{$\cdots$}
\enddata
\tablecomments{(1) IDs of Ly$\alpha$ nebulae sorted by the Ly$\alpha$ isophotal area from the largest to the smallest; (2) field name (3-4) coordinates R.A. and Decl. in equinox with an epoch of J2000; (5-6) narrowband (NB387 or NB400) and g-band magnitude; (6-7) isophotal Ly$\alpha$ area and Ly$\alpha$ luminosity above $2\sigma$ contour in the unit of $\rm arcsec^2$ and $\rm erg~s^{-1}$; (8) projected end-to-end Ly$\alpha$ extent in the unit of physical kpc; (9) spectroscopic redshift if available; (10) comments for the corresponding object (T2: Type II ELAN; R: with radio counterparts; Q: QSO with spectra; IR: IR luminous objects identified by WISE;). Only the first ten sources are listed, and the entire table is available in a machine-readable form in the online journal (now \href{https://www.dropbox.com/scl/fi/pf75tfyok5vvq9p1ufslf/MAMMOTH-Subaru-LyaNebula.txt?rlkey=r3nkfpqw8usr5k9b37lnaexio&st=lkcusg6o&dl=0}{here}).}
\end{deluxetable*}

\subsection{A large sample of Ly\titleAlpha nebulae}
Taking a deep narrowband imaging survey covering an effective area of $\simeq 12 \mathrm{~deg}^2$, we reveal 117 Ly$\alpha$ nebulae at redshift 
$z=2.1-2.3$.
This nebula sample significantly increases the number of known extended Ly$\alpha$ nebulae at this epoch.
These nebulae have Ly$\alpha$ luminosities ranging from $8\times10^{42} \mathrm{~erg~s^{-1}}$ to $4\times10^{44} \mathrm{~erg~s^{-1}}$ and isophotal areas of $14 - 290 \mathrm{~arcsec^2}$.
It is noteworthy that out of 117 nebulae, there are 28 nebulae in our sample with an end-to-end Ly$\alpha$ size over 100 kpc, categorized as ELANe.

In Figure~\ref{fig:labsum}, we show the Ly$\alpha$ luminosities, isophotal areas, and maximum projected sizes of our sample.
For comparison, we overlay results from previous studies \citep{Kikuta2019, Yang2010, ArrigoniBattaia2019, Matsuda2004,Matsuda2011, Borisova2016, Badescu2017, Hennawi2015, Cantalupo2014, Cai2017b}.
However, direct comparisons between studies should be made cautiously due to different selection methods, sensitivity, and redshifts between surveys.
For example, the isophotal areas and sizes were measured using different narrowband filters and luminosity limits.
Therefore, we refrain from over-interpreting the comparisons. 

We find that there is a field-to-field variation in nebula number densities for our eight fields, as no significant difference exists between survey depth and survey area.
This variation can indicate strong clustering of the nebulae, as analyzed by \citet{Zhang2023a} and in our future work.


\subsection{Ivory Nebula: Discovery of an intergalactic nebula}\label{sec:ivory}

\begin{figure*}[ht]
\centering
\includegraphics[width=0.95\linewidth]{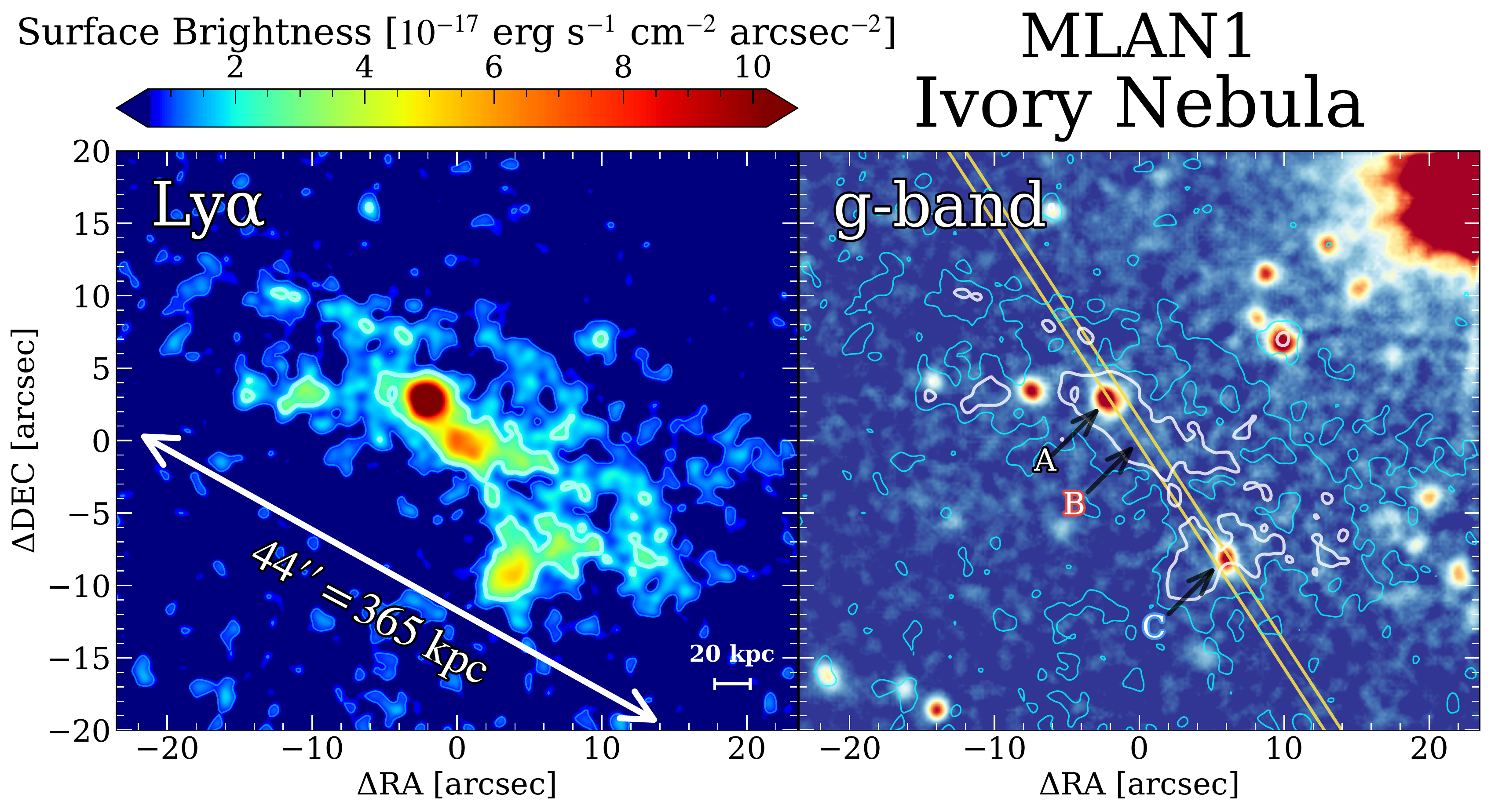}
\includegraphics[width=0.55\linewidth]{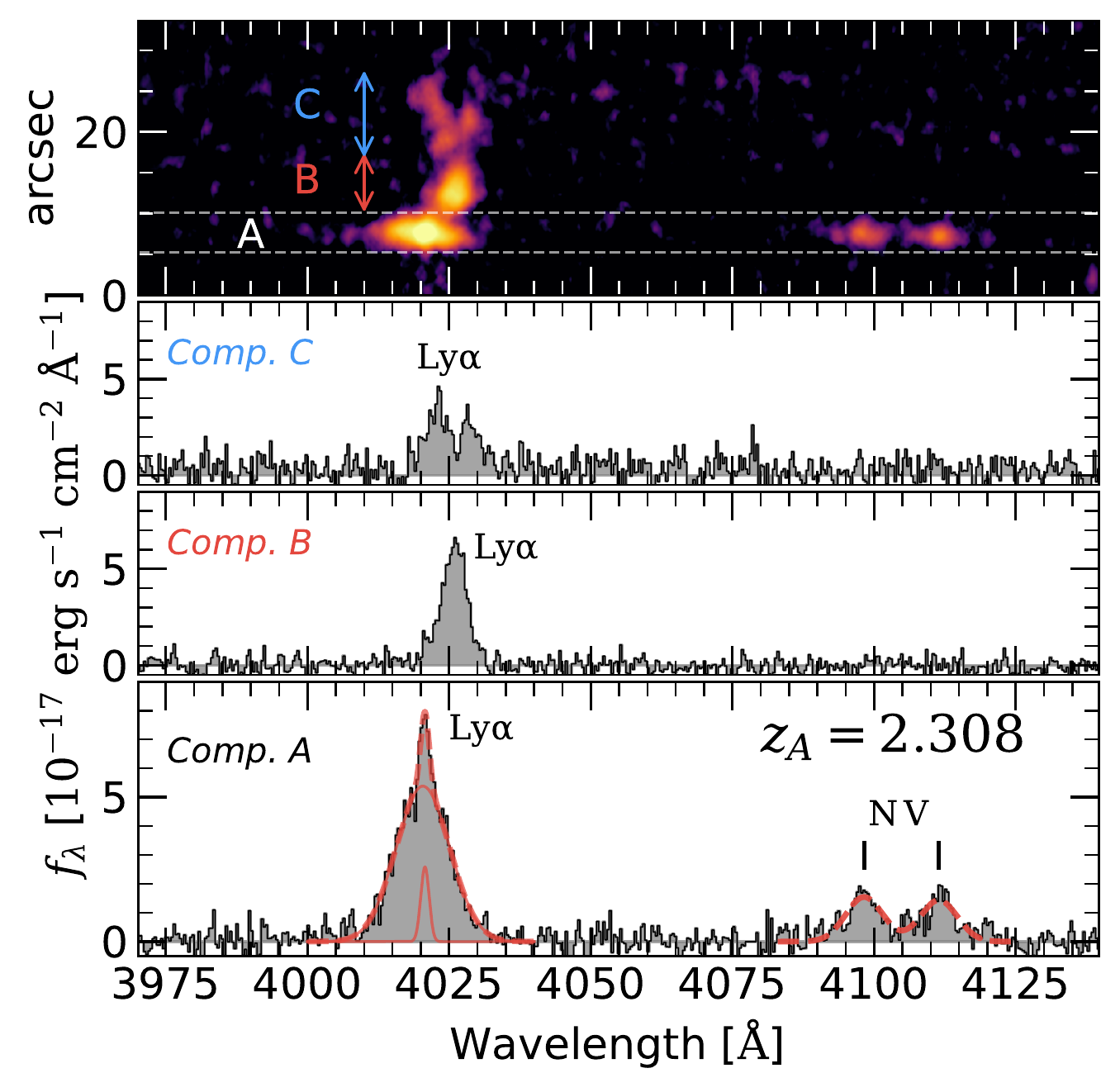}
\includegraphics[width=0.44\linewidth]{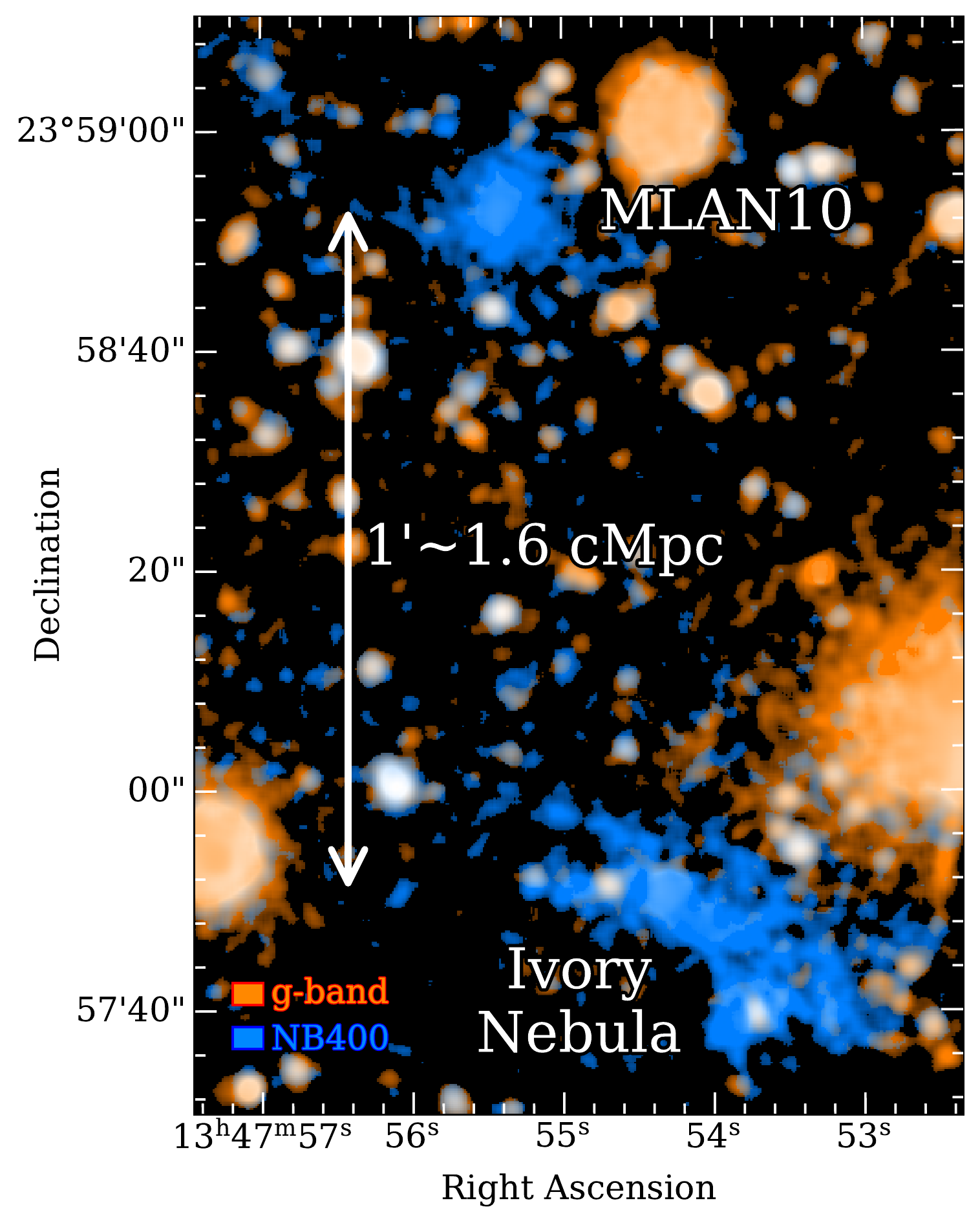}
\caption{\textit{\textbf{Upper Panels:}} Ly$\alpha$ line map (left) and g-band image (right) of the field around Ivory Nebula. The line map reveals the extremely extended Ly$\alpha$ emission of the enormous Ly$\alpha$ nebula. The $2 \sigma$ surface brightness limit is $7.76 \times 10^{-18}~\mathrm{erg} ~\mathrm{s}^{-1}~\mathrm{cm}^{-2}~\mathrm{arcsec}^{-2}$. Cyan and white lines represent the $2\sigma$ and $5\sigma$ Ly$\alpha$ contours, respectively.
Yellow lines denote the long slit to take spectra. 
\textit{\textbf{Lower Left Panel:}} 2D and 1D spectra of the Ivory Nebula. The 2D spectrum shows the spatial extent of Ly$\alpha$ emission along the slit (oriented along the yellow lines in the upper right). 1D spectra were extracted using fixed-width apertures corresponding to the three components labeled (component A, B, and C). The spectra reveal complicated Ly$\alpha$ line profiles and the detection of \ion{N}{5} doublet emission in Component A.
\textit{\textbf{Lower Right Panel:}} Composite RGB image (g-band in red and NB400 in blue) of the ELAN pair composed of Ivory Nebula and MLAN10. Pixels with values of $<2\sigma$ RMS are set in black color. The projected angular separation between these two ELANe is about 1 arcmin.
\label{fig:ivory}}
\end{figure*}

We report the discovery of MLAN1, the largest nebula in our sample.
To emphasize its exceptional properties, we have given this nebula the nickname ``Ivory Nebula''.
The Ly$\alpha$ line map for Ivory Nebula shows extended filamentary structure and well-resolved clumpy Ly$\alpha$ emission along the direction of two galaxies (see Figure~\ref{fig:ivory}).
Ivory Nebula has a Ly$\alpha$ luminosity of $L_{\mathrm{Ly\alpha}} = 2.5\times10^{44} \mathrm{~erg~s^{-1}}$ and an end-to-end angular extent of $D \simeq 44.4\arcsec  \simeq 365 {~\rm kpc}$.
The extreme properties make it one of the largest and most luminous intergalactic nebula discovered in the Universe.

We obtained a long-slit spectrum for the Ivory Nebula along the axis connecting its two associated galaxies (indicated by the yellow lines in the right panel of Figure~\ref{fig:ivory}).
The spectra, zoomed in on the wavelength region surrounding the Ly$\alpha$ emission line, are shown in Figure~\ref{fig:ivory}.
We detect and confirm the presence of spatially extended Ly$\alpha$ emission from this spectacular nebula.
Strikingly, the 2D spectra reveal complex kinematics of the Ly$\alpha$ emitting gas.
Based on the peaks in the Ly$\alpha$ line flux map, we divide the nebula into three components (A, B, and C, as indicated in Figure~\ref{fig:ivory}) along the slit.

Component A is a galaxy with a g-band magnitude of $\mathrm{m_{g}}=24.08$.
In addition to Ly$\alpha$, the spectra show \ion{N}{5} $\lambda\lambda1238.8,1242.8$ doublet emission.
To make the highly ionized \ion{N}{5} emission line possible, $\rm N^{4+}$ with $h\nu > 77.5 \rm~eV$ is required, indicating the presence of an AGN in the component A.
The Ly$\alpha$ line profile of A consists of a broad component with FWHM = 831 km s$^{-1}$  and a narrow Ly$\alpha$ spike with FWHM = 130 km s$^{-1}$, shown in Figure~\ref{fig:ivory}.
Component B can be a diffuse Ly$\alpha$-emitting gaseous ``dark cloud" \citep[\eg,][]{Cantalupo2012} without a UV continuum counterpart.
Component C is another faint continuum source with a g-band magnitude of $\mathrm{m_{g}}=24.36\pm0.06$, which is surrounded by a luminous Ly$\alpha$-emitting cloud with a complicated double-peak feature.
It is noteworthy that Ly$\alpha$ emission peak offsets from the component C about 2.5\arcsec.

Besides Ly$\alpha$, the extended \ion{C}{4} and \ion{He}{2} emissions are detected by our observations from the Keck/Cosmic Web Imager (KCWI), indicating the existence of metal-enriched gas on the CGM scale.
The multi-wavelength observations and a detailed analysis of Ivory Nebula will be posted in our upcoming paper (Li et al. in prep).

Ivory Nebula is located in the large-scale structure of an LAE overdensity.
This structure is confirmed as an overdensity with 12 nebulae (eight bright nebulae with Ly$\alpha$ luminosity larger than $2.5\times10^{43}\mathrm{~erg~s^{-1}}$) in the volume of $\simeq 40\times20\times90$ cMpc$^3$ \citep{Zhang2024}.
The bright nebula volume density is comparable and even higher than the well-known protocluster SSA22 \citep[\eg,][]{Steidel2000, Matsuda2004}.

We also want to emphasize the remarkably close ELAN pair within this overdensity, comprising the Ivory Nebula and MLAN10 as shown in Figure~\ref{fig:ivory}.
MLAN10 lies approximately 1 arcmin ($\simeq 500~\mathrm{kpc}$ or $1.6~\mathrm{cMpc}$ at $z=2.3$) away from the Ivory Nebula with Ly$\alpha$ luminosity of $8.1\times10^{43}\mathrm{~erg~s^{-1}}$ and Ly$\alpha$ extent of $ 120.8 \mathrm{~kpc}$.
We note that MLAN10 is almost located in the projected line along component A and C of the Ivory Nebula, indicating ELANe as nodes of the filamentary cosmic web.
We expect that deeper observation can reveal the cosmic-web filaments \citep{Martin2023, Umehata2019} probed by relatively bright Ly$\alpha$ emission bridge (\eg, $\mathrm{SB_{Ly\alpha}} > 10^{-18} \mathrm{~erg~s^{-1}~cm^{-2}~arcsec^{-2}}$) and embedded dwarf galaxies connecting the two ELANe.

\section{Diverse Populations of Ly\titleAlpha Nebulae}
Our sample presents 117 Ly$\alpha$ nebulae spanning a wide range of properties, \eg, luminosities and sizes.
In order to investigate properties of galaxies associated with these nebulae, we cross-match our nebula sample with multi-wavelength observations, which ultimately showcase a diversity of nebulae.
In this section, we explore this diversity focusing on multiple diagnostics, including UV fluxes, infrared (IR) fluxes, radio emissions, and AGN activity.

\subsection{Type II nebulae missing UV-bright sources}
\label{sec:uv}

\begin{figure*}[th]
\centering
\includegraphics[width=\linewidth]{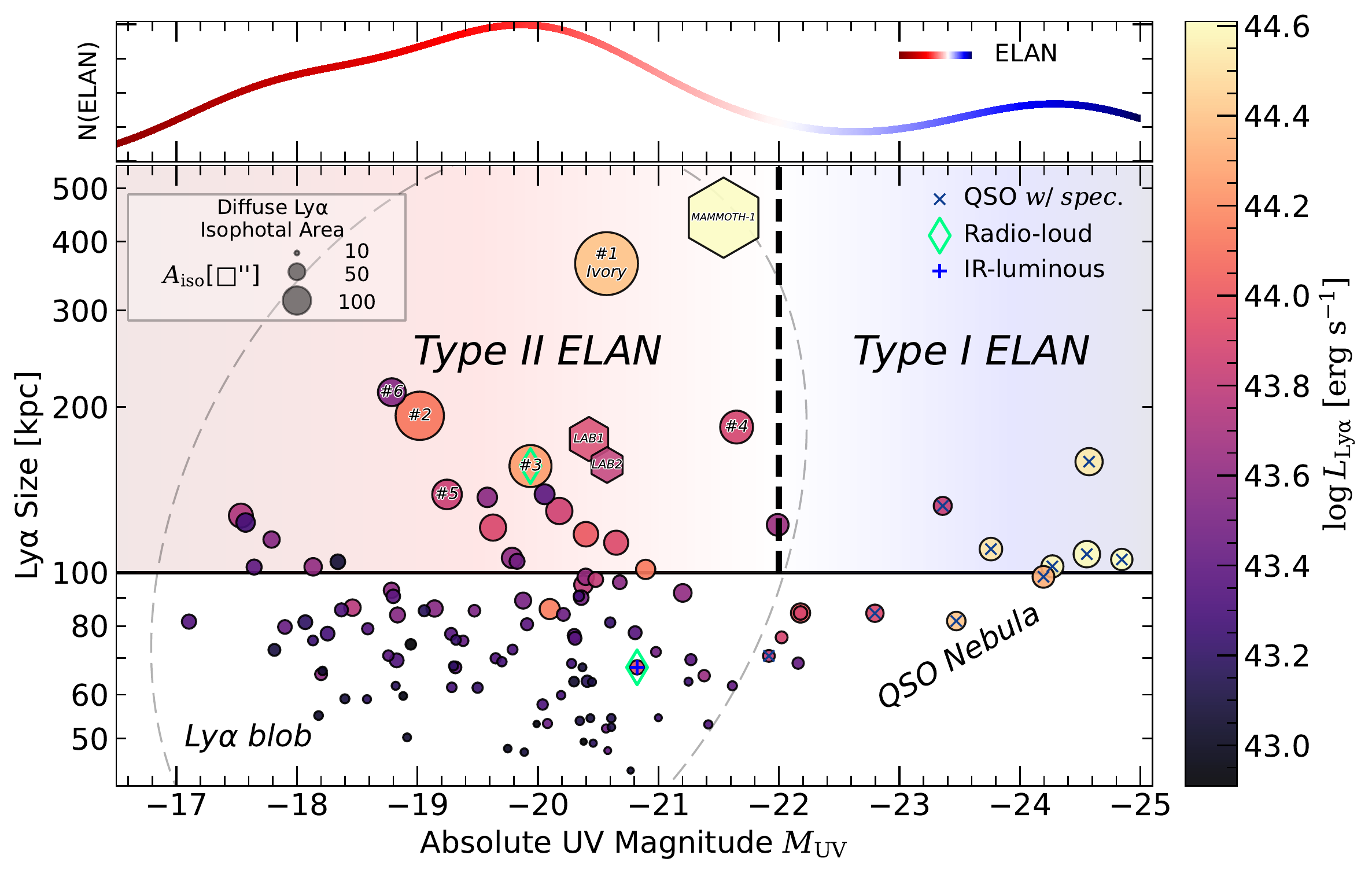}
\caption{
Diagnosis of Type II ELANe, enormous Ly$\alpha$ nebulae ($>100~\rm kpc$) with UV-faint ($M_\mathrm{UV}>-22$) host galaxies. Type II ELANe are the majority ($\sim 80\%$) in the population of enormous Ly$\alpha$ nebulae. The x-axis is the rest-frame absolute UV magnitude. The y-axis denotes the projected Ly$\alpha$ angular size. The size of each circle denotes the diffuse Ly$\alpha$ isophotal area. The color of each circle denotes the Ly$\alpha$ luminosity. The UV magnitude distribution of ELANe is shown in the upper panel with a bimodal feature. Type II ELANe are located in the upper left corner with a faint UV continuum but widespread Ly$\alpha$ emissions, while Type I ELANe are located at the upper right of this figure with a strong UV continuum. The background of this figure is colored just for visual distinction. The largest six nebulae in this work are labeled, which all are Type II nebulae. For summary and comparison, a few Type II nebulae reported in previous literature are drawn as hexagons in this figure, including MAMMOTH-1 \citep{Cai2017b, Zhang2023c}, SSA22-LAB1, SSA22-LAB2 \citep{Steidel2000, Matsuda2004}. QSO confirmed by spectroscopic observations (blue cross), radio galaxies (green diamond), and IR-luminous DOGs (blue plus) are overlaid on the figure.
\label{fig:uv-faint}}
\end{figure*}

\begin{figure}[ht]
\centering
\includegraphics[width=\linewidth]{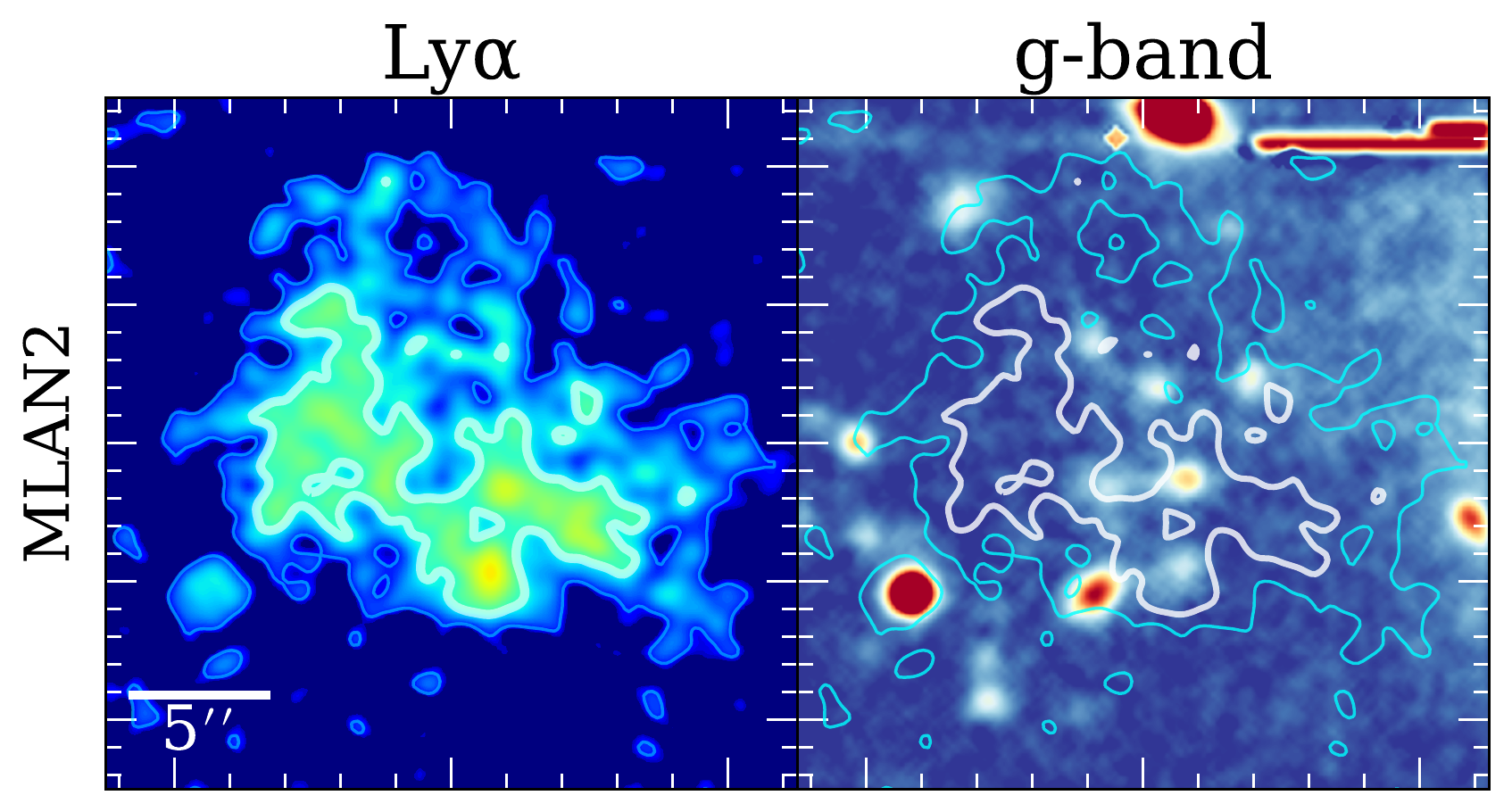}
\includegraphics[width=\linewidth]{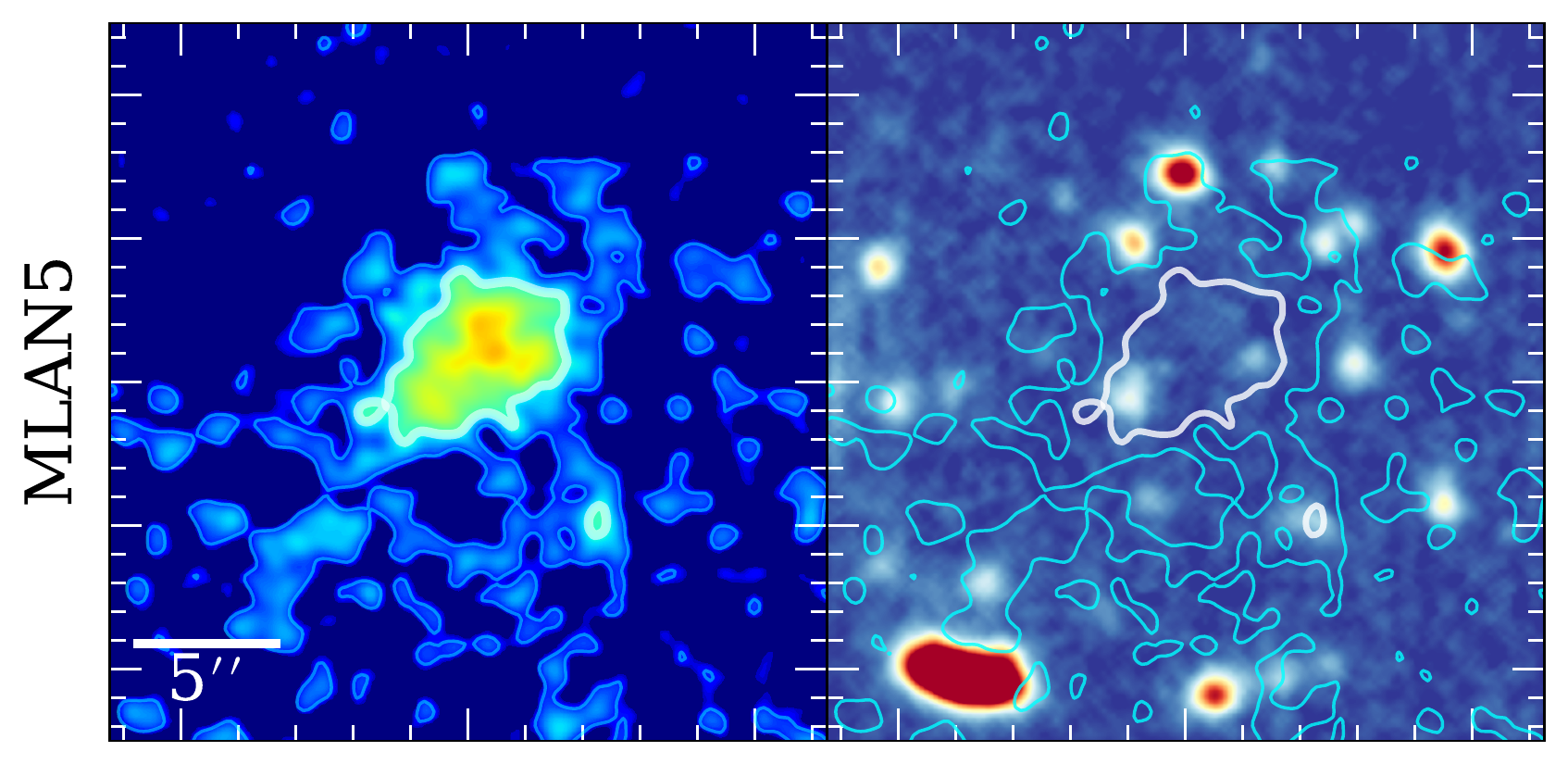}
\caption{Two representative type II ELANe. The left panels present the smoothed continuum-subtracted Ly$\alpha$ line maps, and the right panels show the HSC g-band imaging (rest-frame UV). The cyan and white lines represent the 2$\sigma$ and 5$\sigma$ Ly$\alpha$ surface brightness contours. These enormous nebulae are associated with UV-faint sources.
\label{fig:ghost-LAB}}
\end{figure}

In normal circumstances, the luminous and extended Ly$\alpha$ emission can be explained by the fluorescence of embedded energetic UV sources \citep[\eg,][]{Cantalupo2005, Kollmeier2010, Overzier2013}, including starburst and QSOs.
However, we find that the majority of enormous Ly$\alpha$ nebulae are associated with UV-faint sources.
We categorize this intriguing class of enormous Ly$\alpha$ nebulae as Type II ELAN, defined empirically as enormous Ly$\alpha$ nebulae ($\gtrsim100~\rm kpc$) lacking any UV-bright source ($M_{\rm UV}\gtrsim-22$).
Oppositely, we define enormous Ly$\alpha$ nebulae with UV-bright sources ($M_{\rm UV}\lesssim-22$) as Type I ELANe.
The boundary of absolute UV magnitude ($M_{\rm UV}=-22$) used to differentiate between two types is determined empirically by identifying the critical luminosity where the UV luminosity functions of QSOs and galaxies at $z\sim2$ intersect \citep[e.g.,][]{Shen2020}.
The galaxy dominates the faint UV population in $M_{\rm UV}>-22$ while the QSO dominates the bright UV population in $M_{\rm UV}>-22$.
Because of the the faintness of their UV continuum, it is extremely challenging to reveal Type II ELANe through current QSO and galaxy surveys.
In summary, we define two types of enormous Ly$\alpha$ nebulae based on their associated UV sources:
\begin{itemize}
    \item Type I: These nebulae have associated UV-bright sources ($M_{\rm UV}\gtrsim-22$), typically QSOs.
    \item Type II: These nebulae have no detectable UV-bright sources ($M_{\rm UV}<-22$), typically galaxies, despite their extended and bright Ly$\alpha$ emission.
\end{itemize}


In our sample, 28 out of 117 ($\simeq 24\%$) nebulae can be identified as enormous Ly$\alpha$ nebulae with projected size over than $100~\rm kpc$.
Strikingly, 22 of these 28 ELANe ($\simeq 78.6\%$) satisfy the criteria of Type II ELANe, \ie, enormous nebulae missing associated bright UV continuum sources, including the Ivory nebula.
It is noteworthy that the Ivory Nebula satisfies the criteria of Type II ELAN, with the UV continuum of component A has been observed with $M_\mathrm{UV}=-20.5$ in deep g-band imaging (refer to \S\ref{sec:ivory} and Fig.~\ref{fig:ivory}).
We estimate UV magnitude from the continuum flux density following the equation of $M_{UV} = -2.5\log(f_{\nu, \rm cont}) - 48.6 -5\log(D_L/10\rm~pc) + 2.5\log(1+z)$, where $f_{\nu, \rm cont}$ is continuum flux density estimated as shown in \S~\ref{sec:nebula-selection} and $D_L$ is the luminosity distance.
We regard the brightest target overlapped with the nebula as the associated galaxy except spectroscopic targets. 
Figure~\ref{fig:uv-faint} shows the diagram comparing Type II ELAN against other classes including Type I ELAN, QSO nebula, and Ly$\alpha$ blob.
We find there is a bimodal distribution of UV magnitude for all ELANe, with Type I ELANe (QSO nebulae) exhibiting bright UV continua and Type II ELANe associated with faint UV fluxes.
Note that the UV magnitude valley of this bimodal distribution coincide with the boundary magnitude ($M_{\rm UV}=-22$) used to differentiate between two types.
Our sample shows clearly that the ELAN sample is dominated by Type II with a fraction of $\sim78.6\%$.
Therefore, previous ELAN searches mainly on pre-selected QSOs are highly incomplete \citep[\eg,][]{Borisova2016, ArrigoniBattaia2019}.

Besides Ivory Nebula as a typical Type II, we present two other representative Type II nebulae, MLAN2 and MLAN5, in Figure~\ref{fig:ghost-LAB}.
MLAN2 has an extended Ly$\alpha$ emission out to 192.9 kpc with a Ly$\alpha$ luminosity of $1.32 \times 10^{44}\mathrm{~erg~s^{-1}}$.
Its morphology is nearly circular, with two brighter lobes.
MLAN2 exhibits a flat Ly$\alpha$ surface brightness profile of $\simeq 1 - 5 \times 10^{-17}\mathrm{~erg~s^{-1}~cm^{-2}~arcsec^{-2}}$.
MLAN5 shows 138.8 kpc filamentary Ly$\alpha$ emission with a luminosity of $6.85 \times 10^{43}\mathrm{~erg~s^{-1}}$.
The Ly$\alpha$ emission has an emitting core with multiple flux peaks.
Despite enormous Ly$\alpha$ structures, no bright UV continuum is detected associated with the nebulae.

The existence of a considerable sample of Type II ELANe poses a puzzle since their large Ly$\alpha$ extent and copious Ly$\alpha$ luminosities could be difficult to explain without copious UV photons.
Consequently, additional Ly$\alpha$ powering mechanisms (\eg, gravitational cooling) may play an important role in this category of sources.
In fact, the very first discovered blobs \citep[\eg, SSA22 LAB1 and LAB2 in][]{Steidel2000} and MAMMATH-1 nebula\citep{Cai2017b} are all Type II ELANe (shown in Fig.~\ref{fig:uv-faint}).
The dominance of Type II among ELAN samples shown by our work further highlights their importance and demonstrates the universality of these mechanisms.
Here we summarize and propose several possible mechanisms to explain these Type II ELANe.

(1) Obscured AGN.
The enormous nebulae are illuminated by heavily obscured AGNs.
For example, obscured AGNs have been found in the MAMMOTH-1 nebula and SSA22-LAB2, from the X-ray \citep[\eg,][]{Geach2009,Zhang2023b} and IR observations\citep[\eg,][]{ArrigoniBattaia2018b}.
It is noteworthy that these obscured AGNs are also called Type II AGNs, which can be a possible scenario for powering the Type II ELANe defined in this work.
However, it is not expected that each Type II ELAN hosts a Type II AGN.
We use the term ``Type II'' here to emphasize the UV-faint properties that can be caused by the dusty nature.
In \S~\ref{sec:ir}, we discuss the mid-infrared properties of our nebulae related to this obscured AGN scenario and corresponding observations in more detail.
It is noteworthy that these obscured AGNs are always hosted by dusty starburst galaxies, as shown in the following paragraph.

(2) Dusty starburst.
The nebulae may be powered by a dusty starburst that is not detectable in the rest-frame UV due to its dusty nature \citep[\eg,][]{Cen2013}.
In this scenario, infrared/sub-mm observations using JWST and ALMA could potentially reveal luminous hidden powering sources within the nebulae.
For example, SSA22-LAB1 has been confirmed to be composed of numerous sub-millimeter galaxies (SMGs) with violent star formation\citep[\eg,][]{Umehata2021}.
It is noteworthy that, for most of Type II ELANe, the Ly$\alpha$ flux peak tends to displaced from the flux peak of the galaxy UV continuum (see Fig.~\ref{fig:ivory} and Fig.~\ref{fig:ghost-LAB}).
This phenomenon has been observed before \citep[\eg,][]{Dey2005} and can be reproduced by starburst model \citep[\eg,][]{Cen2013} by combining complicated effects of galaxy clustering, faint under-resolved sources, and dust.
Our observations indicate a tentative ubiquity of this Ly$\alpha$ peak displacement for Type II ELANe.

(3) Gravitational cooling.
The Ly$\alpha$ emission is due to the cold accretion cooling radiation of filamentary gas onto a massive dark matter halo \citep[\eg,][]{Dijkstra2009,Faucher-Giguere2010, Haiman2000, Rosdahl2012, Daddi2021}.
Sensitive sub-millimeter observations could detect cold molecular gas within the halo that may fuel the star formation in the galaxy \citep{Ao2020}. 
The cooling flow can contribute to the Ly$\alpha$ at the ongoing stage of massive galaxy formation when there has been a massive dark matter halo in the center \citep{ArrigoniBattaia2018b, Zhang2023c}.

(4) The relic emission.
In this case, the nebula is a relic structure - the UV photons required to power it and Ly$\alpha$ photons were emitted in the past when the central source was still active.
The Ly$\alpha$ photons can continue scattering in the nebula \citep[\eg,][]{Dijkstra2009,Hayes2011}, even after the central engine shuts off.
The most likely central engines powering the original Ly$\alpha$ emission in this scenario are AGNs with rapid duty cycles (AGN flickering), which can emit copious ionizing radiation that gets converted to Ly$\alpha$ through recombination \citep[\eg,][]{Schirmer2016}.
Resonant Ly$\alpha$ photons are efficiently stored and will be gradually released, decorrelating from AGN variability on scales of up to $10^6$ years depending on the optical depth.
After the central engine shuts off, the nebula can continue glowing in Ly$\alpha$ emission for an extended period due to recombination equilibrium.

(5) Numerous faint interacting galaxies.
The nebulae are powered by numerous interacting sources that fall below the detection threshold.
A cluster of faint star-forming galaxies or lower-luminosity AGN could fuel the Ly$\alpha$ fluorescence.
The existence of such faint UV sources that cannot be detected individually has been shown through stacking of LAEs \citep[\eg,][]{Kikuta2023}, and it has been demonstrated that they contribute significantly to the extended Ly$\alpha$ emissions.
Further sensitive observations could potentially uncover these hidden power sources within the individual nebulae.

The categorization of Type I and Type II ELANe could be influenced by how UV sources are identified and UV flux measurements are conducted.
One factor to consider is the spatial offset that could exist between the Ly$\alpha$ peak and the UV sources.
Typically, only the brightest target within the nebula is taken into account in this work.
Consequently, the UV fluxes may appear fainter than they actually are.
Moreover, certain emission lines like \ion{C}{4} can enhance the estimate of UV magnitude based on broad-band photometry, leading to measured UV fluxes that appear brighter than they truly are.
Despite these limitations, it is significant that the Type IIs occupy the majority of ELANe.

Type II nebulae represent the most interesting and most important categories of the diverse Ly$\alpha$ nebula population.
Note that with only deep NB387/NB400 and g-band imaging observations, detailed mechanisms of these Type II nebulae in our sample are hard to determine.
However, the fact that the majority of the Ly$\alpha$ nebula sample is Type II has been clearly shown.
Further multi-wavelength follow-up, especially infrared, (sub)mm, and X-ray (\eg, using JWST, ALMA, and Chandra), will be essential for unraveling the power sources and mechanisms.
Their study promises to reveal new insights into the interplay between galaxies, supermassive black holes, and CGM in the peak epoch of AGN activity and cosmic star formation.

\subsection{Infrared properties of Ly\titleAlpha nebulae}
\label{sec:ir}

Infrared observations provide key insights into the powering sources of Ly$\alpha$ nebulae, particularly those lacking bright UV/optical counterparts (Type II).
Obscured star formation and AGN activity can be revealed in the IR wavelength \citep[\eg,][]{Lyu2022}, which could play an important role in powering extended Ly$\alpha$ emission.
In this section, we investigate the mid-infrared properties of our sample.

To analyze the infrared properties of our nebula sample, we cross-matched our 117 nebulae with observations of Wide-field Infrared Survey Explorer \citep[WISE;][]{Wright2010}.
WISE surveyed the entire sky at wavelengths of 3.4, 4.6, 12, and 22 $\mu$m down to 5$\sigma$ depths of 0.08, 0.11, 1, and 6 mJy, respectively.
For our nebulae at $z\approx2$, these four WISE bands cover the 1, 1.5, 4, and 7 $\mu$m rest-frame wavelength.

The crossmatch results showed WISE detection for 25 out of 117 nebulae ($\approx$21\%) in at least one band.
All these 25 objects are detected in the W1 (3.4 $\mu$m) band, with 20, 11, and 4 objects detected in the W2 (4.6 $\mu$m), W3 (12 $\mu$m), W4 (22 $\mu$m) bands respectively.
The WISE-detected counterparts of Ly$\alpha$ nebulae have typical AB magnitudes of W1, W2 $>17$.
Note that we show all WISE photometry in AB magnitude, which is converted from the Vega magnitude following the relation: $\rm (W1,W2,W3,W4)_{AB} = (W1,W2,W3,W4)_{Vega} + (2.699, 3.339, 5.174, 6.620)$.

\begin{figure}
\centering
\includegraphics[width=\linewidth]{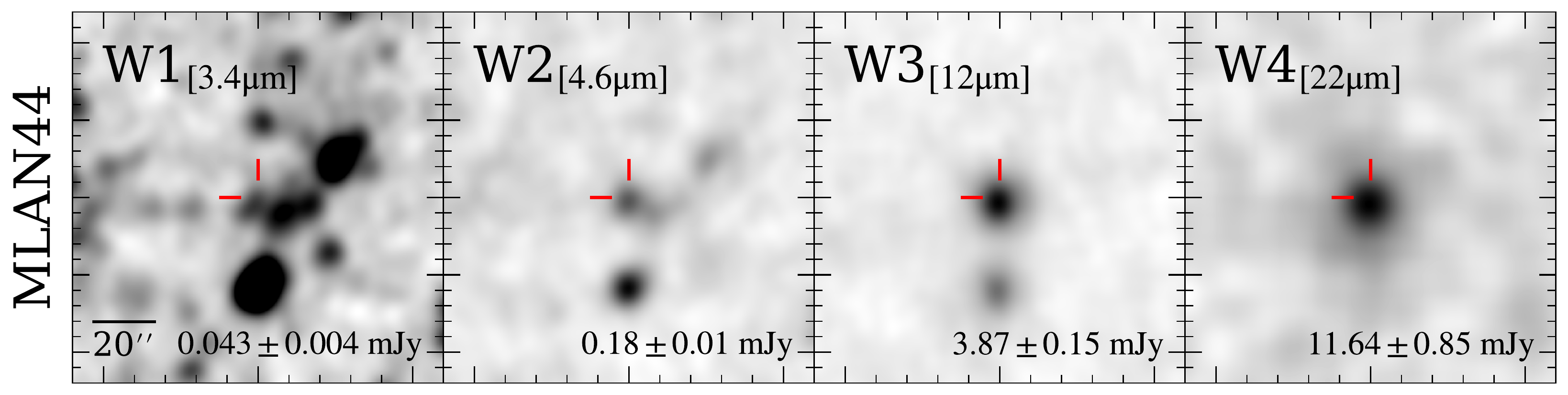}
\includegraphics[width=\linewidth]{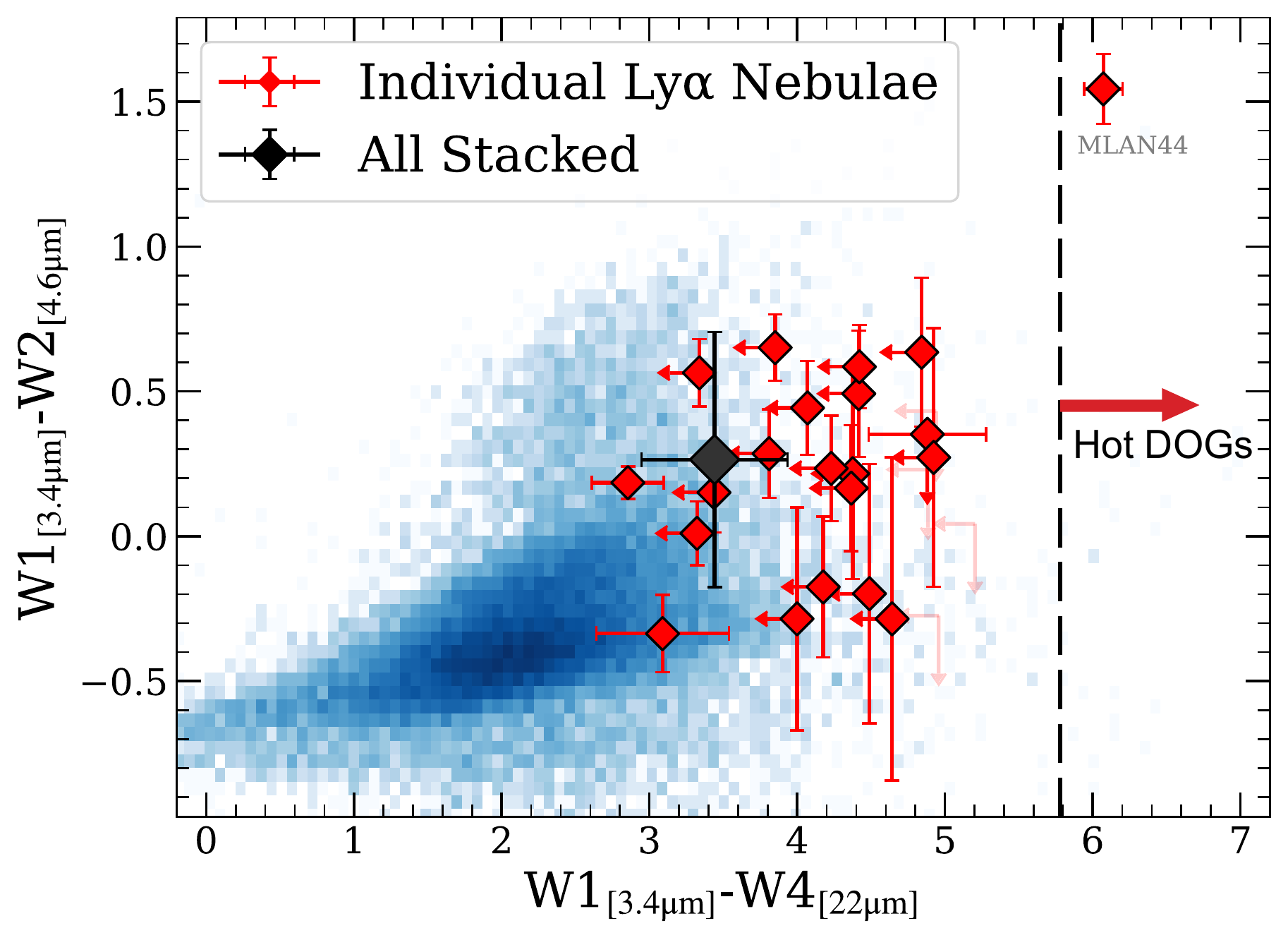}
\caption{Upper: Cutout images of MLAN44 in the WISE W1 (3.4$\mu$m), W2 (4.6$\mu$m), W3 (12$\mu$m), and W4 (22$\mu$m) bands, from left to right. The cutout size is $2' \times 2'$.
Lower: The WISE infrared color-color diagram of the matched and stacked sources in our sample. The blue-indexed 2D histogram denotes the distribution of sources with $< 0.3$ mag errors in W1 (3.4$~\mu$m), W2 (4.6$~\mu$m), and W4 (22$~\mu$m) in the all-sky release catalog with Galactic latitude $b>40^\circ$. Objects to the right of the dashed line are hot dust-obscured galaxies \citep{Eisenhardt2012}.
\label{fig:wise-color}}
\end{figure}

Among the WISE-detected nebulae, one target (MLAN44) displays an extremely bright flux in the W3 and W4 bands (Fig. \ref{fig:wise-color}).
The WISE magnitudes of MLAN44 are $m_\mathrm{3.4\mu m} = 17.10\pm0.10$, $m_\mathrm{4.6\mu m} =14.92\pm0.06$, $m_\mathrm{12\mu m}=9.76\pm 0.04$, and $m_\mathrm{22\mu m}=7.12\pm0.08$ in Vega magnitude. 
MLAN44 is the most mid-IR luminous source in our sample, despite having a faint UV counterpart ($m_\mathrm{g} = 23.76\pm0.02$).
The extended Ly$\alpha$ emission around MLAN44 reaches 67 kpc in projection with a luminosity of $L_\mathrm{Ly\alpha} = 8.69 \times 10^{43} \mathrm{~erg~s^{-1}}$.
The extreme infrared colors and luminosity, combined with the spatial extent and high Ly$\alpha$ luminosity, suggest that MLAN44 hosts an obscured AGN or dusty starburst that likely powers the Ly$\alpha$ emission through photo-ionization.

MLAN44 represents one case of the luminous infrared galaxy (LIRGs), a class of objects identified by their high infrared luminosity and faint optical counterparts \citep[\eg,][]{Dey2008, Wu2012, Eisenhardt2012, Liu2022}.
In nature, these galaxies are explained by the dust-obscured violent starburst called dust-obscured galaxy (DOG) or obscured AGN activity (hot DOG or Type-II QSO).
In this work, we do not make specific distinctions, and refer to them collectively as (hot) DOGs.
The WISE colors and luminosities of MLAN44 are consistent with known hot DOGs ($\rm W1-W4\lesssim5.8$) at similar redshifts \citep[\eg,][]{Ishikawa2023}.
Mid-IR imaging for hot dust and spectroscopy targeting polycyclic aromatic hydrocarbon (PAH) emission features can help distinguish the contributions from AGN activity and star formation\citep{Colbert2011}.
The JWST MIRI observations are expected for this population of Ly$\alpha$ nebulae to unravel the dusty nature and examine the powering mechanism of the source.

Despite the obscured nature, almost all hot DOGs have Ly$\alpha$ emission and are detected as Ly$\alpha$ emitters \citep{Bridge2013}.
Furthermore, a significant fraction of hot DOGs exhibit Ly$\alpha$ nebulae extending to scales of $>30$ kpc, well beyond the dust envelopes \citep[\eg,][]{Bridge2013, Dey2005}.
However, with the detection limit of WISE, only the most luminous infrared galaxies can be detected.
The fraction of nebulae with WISE-detected luminous IR counterparts can be estimated to be $1/117 \sim 0.9\%$ with a number density of $1.14\times10^{-7}\mathrm{~cMpc^{-3}}$.
This indicates the extremely low number density of Ly$\alpha$ nebulae with hot DOGs.
More sensitive mid-IR and Far-IR/sub-millimeter observations can better reveal the dust-obscured nature.


\subsection{Ly\titleAlpha Nebulae associated with Radio Galaxies}
\label{sec:radio}
\begin{figure}[t]
\centering
\includegraphics[width=\linewidth]{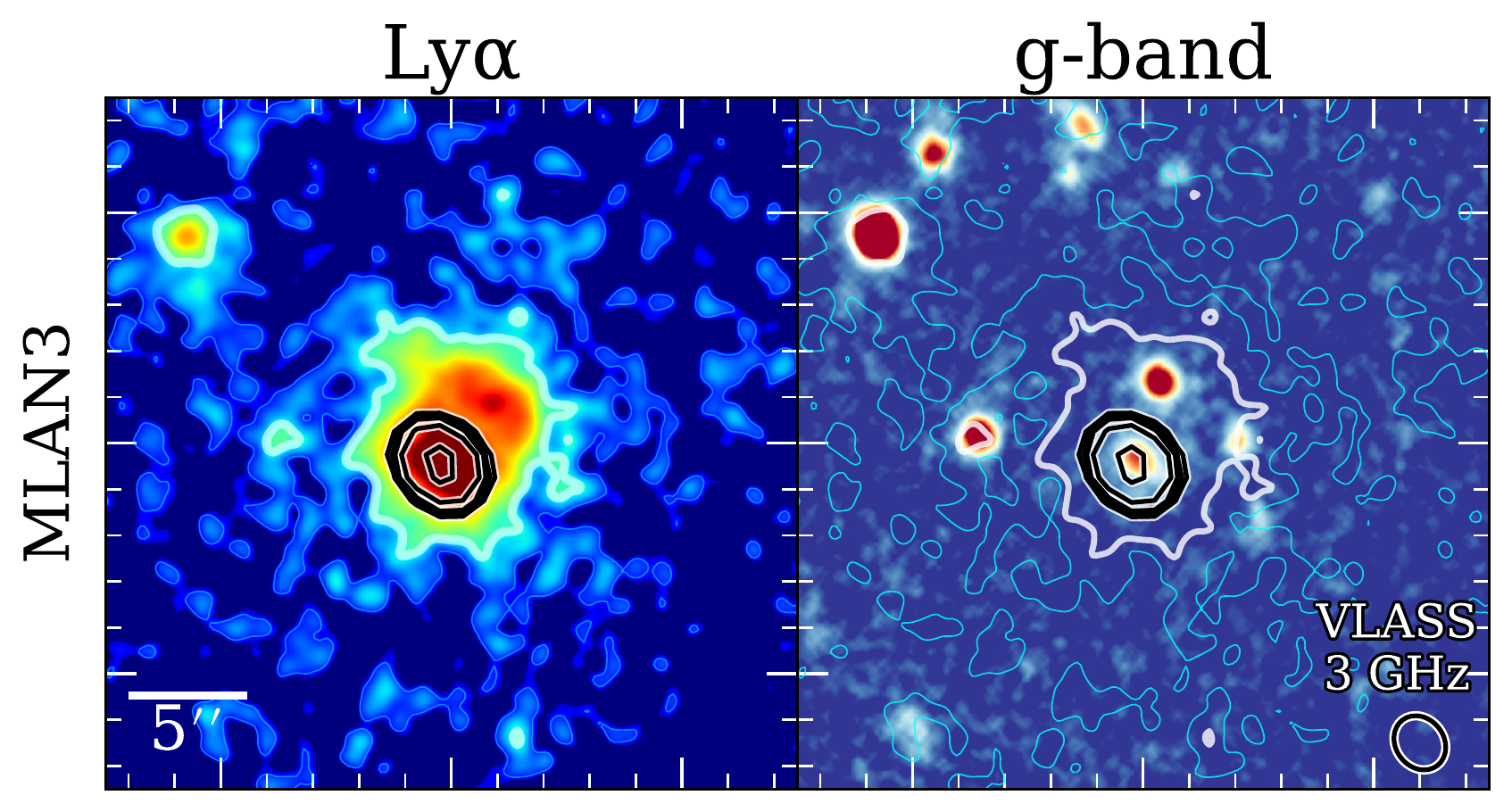}
\includegraphics[width=\linewidth]{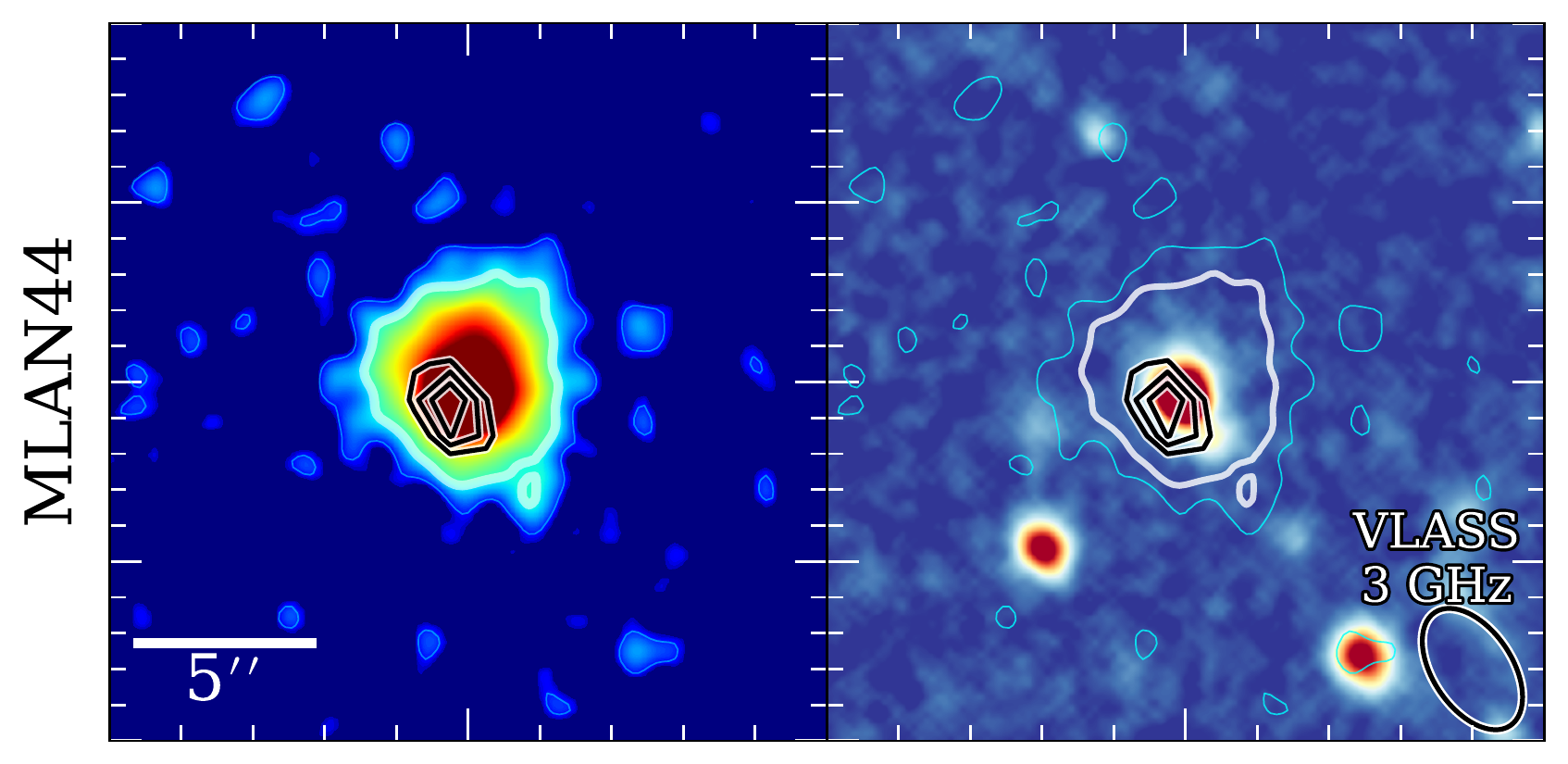}
\caption{Two Ly$\alpha$ nebulae with radio counterparts. The left panels present the smoothed continuum-subtracted Ly$\alpha$ line maps, and the right panels show the HSC g-band imaging. The cyan and white lines represent the 2$\sigma$ and 5$\sigma$ Ly$\alpha$ surface brightness contours. VLA 3 GHz radio observations from VLASS are overlaid as black contours of $5,6,7,10,20\sigma$ where $\sigma = 0.15\rm~mJy$.
\label{fig:radio-LAB}}
\end{figure}

The first extended Ly$\alpha$ nebulae were discovered around HzRGs and radio quasars \citep[\eg,][]{McCarthy1987}, highlighting the diversity of radio properties early on.
The interaction of relativistic jets from central SMBHs with ambient gas is believed to contribute to powering and distributing Ly$\alpha$ emission on large scales \citep[\eg,][]{McCarthy1993}.
However, owing to the small sample sizes of Lya nebulae and the fact that powerful radio sources are found only in a very small fraction of high-$z$ galaxies, very few Lya nebulae have radio counterparts.
In this section, we compile the public radio survey to quantify the radio properties of our nebula sample.

We match our 117 nebulae with public data archives of radio observations, including the VLA Faint Images of the Radio Sky at Twenty-Centimeters (FIRST) Survey \citep{White1997, Helfand2015}, the NRAO VLA Sky Survey \citep[NVSS,][]{Condon1998}, VLA Sky Survey \citep[VLASS, \eg,][]{Lacy2020}, and TIFR GMRT Sky Survey \citep[TGSS,][]{Intema2017}.
These four surveys cover all eight fields in our work.
With the matching radius of 10\arcsec, two nebulae (MLAN3 and MLAN44, shown in Figure~\ref{fig:radio-LAB}) have radio counterparts in the imaging data of these four surveys.

The VLASS, FIRST/NVSS, and TGSS surveys capture the radio emission at 3.0 GHz, 1.4 GHz, and 150 MHz, respectively.
At a depth of FIRST (flux limits of $\sim$1 mJy), the 115 non-detections place upper limits on the 1.4 GHz radio luminosities of our sample at $\lesssim4\times10^{25} \rm~W~Hz^{-1}$.
For the two nebulae with radio detections, MLAN3 has a 1.4 GHz peak flux density of $10.85\pm0.15$ mJy in FIRST and $11.9\pm0.5$ mJy in NVSS.
MLAN44 has a lower 1.4 GHz peak flux of $2.71\pm0.13$ mJy in FIRST and $4.5\pm0.5$ mJy in NVSS.
At 3 GHz, VLASS measures flux densities of $4.62\pm0.14$ mJy and $1.30\pm0.13$ mJy for MLAN3 and MLAN44, respectively.
In the lower frequency TGSS 150 MHz first alternative data release (ADR1), the sources have peak fluxes of $55.3\pm6.8$ mJy (MLAN3) and $25.9\pm4.3$ mJy (MLAN44).
Combining these multi-frequency measurements, we can derive radio spectral indices.
Adapting values from \citet{deGasperin2018} that matched sensitivity and resolution across the surveys, the best-fit spectral indices are $-0.80\pm0.09$ for MLAN3 and $-0.81\pm0.17$ for MLAN44. 

As revealed by previous studies \citep[\eg,][]{Villar-Martin2007b, Shukla2021, Wang2023b}, Ly$\alpha$ nebulae appear to be nearly ubiquitous around HzRGs, implying a $\sim$100\% Ly$\alpha$ nebulae detection rate for HzRGs.
The prevalence of Ly$\alpha$ nebulae around HzRGs is explained by their gas-rich environments and AGN activity, both of which are conducive to powering and scattering copious Ly$\alpha$ photons on large scales.
The shock-induced radiation powered by radio jets and fluorescent Ly$\alpha$ emission attributed to photo-ionization by central AGNs can dominate for these Ly$\alpha$ nebulae with radio-loud counterparts.

However, the situation changes when turning the argument around.
Only two out of 117 ($\simeq 2\%$) nebulae in our sample have radio counterparts, corresponding to a total volume density of $2.3\times10^{-7}\mathrm{~cMpc^{-3}}$.
The low fraction of the powerful radio galaxies associated with Ly$\alpha$ nebulae implies that radio galaxies are not frequently responsible for the extended luminous nebulae.
Numerous previous LAB surveys \citep[\eg,][]{Matsuda2004, Yang2009, Kikuta2019} have shown that the majority of Ly$\alpha$ nebulae have radio-quiet natures.
In summary, HzRGs can be a subset of the whole Ly$\alpha$ nebula sample.

Despite the rarity of radio-loud Ly$\alpha$ nebulae in our sample, it is noteworthy that we find an excess in the nebula number density for the two fields J0240 and J1349 that host the radio-detected nebulae.
The number densities are $1.70\times10^{-5}\mathrm{~cMpc^{-3}}$ in the J0240 field and $2.60\times10^{-5}\mathrm{~cMpc^{-3}}$ in the J1349 field, a factor of $\sim2.5-4\times$ higher than the average LAB number density at z=2.3 \citep[\eg,][]{Yang2010}.
Note that this nebula density excess is averaged in a vast cosmological volume of $\sim150\times150\times90~\rm cMpc^3$.
This tentatively suggests a correlation between the presence of radio-loud Ly$\alpha$ nebulae and large-scale Ly$\alpha$ nebula overdensities.
On the other hand, radio-loud Ly$\alpha$ nebulae could be related to the existence of ELANe.
Among the 28 ELANe in our sample, 18 ($\sim$64\%) are located in the J0240 and J1349 fields.
Although it is well known that both HzRGs and Ly$\alpha$ nebulae trace the proto-clusters at high redshift \citep[\eg,][]{Mayo2012, Dannerbauer2014, Matsuda2004, Venemans2007, Yang2010, Badescu2017, Zhang2024}, not all galaxy overdensities have ELANe or a large group of Ly$\alpha$ nebulae.
There are some invisible additional parameters affecting the illumination of Ly$\alpha$ nebulae.

One possible invisible parameter is the gas environment of neutral Hydrogen (\ion{H}{1}) in the IGM.
The gas-rich environment could simultaneously promote the young radio galaxy and Ly$\alpha$ nebulae. 
This could lead to the observation that radio galaxies are often found in the centers of assembling protoclusters, as progenitors of local giant central cluster galaxies \citep[\eg,][]{Hatch2009, Wylezalek2013}.
Therefore, it is possible that radio source are predominantly found in the largest Ly$\alpha$ overdensities \citep[\eg,][]{Kurk2000, Pentericci2000}.
Furthermore, Ly$\alpha$ nebulae and ELANe tends to live in these galaxy overdensities \citep[\eg,][]{Steidel2000, Dey2005, Matsuda2005, Prescott2009, Cai2017b, ArrigoniBattaia2018a}.
Therefore, while rare overall, these giant radio-loud Ly$\alpha$ nebulae could be effective tracers of the massive overdensities with a number excess of Ly$\alpha$ nebula and the emergence of ELANe in the large-scale structure.
Further studies of larger samples are still needed to solidify the connection between radio-loud galaxies, nebula overdensities, and protocluster environments.

Except for the all-sky radio survey, we further crossmatch our sample with the Low-Frequency Array (LOFAR) Two-metre Sky Survey (LoTSS), which can detect fainter radio sources at the low-frequency radio band \citep{Shimwell2022}.  
However, the public LoTSS DR2 data release currently only covers two of our fields (J1349 and J0755). 
Four nebulae (MLAN3, MLAN25, MLAN16, and MLAN63) in these two fields have radio counterparts in the LoTSS imaging.  
With the exception of MLAN3, the other three galaxies have relatively faint fluxes ($<1$ mJy at 150 MHz), below the detection limit of the four radio surveys we matched.   

It is noteworthy that LOFAR imaging reveals giant radio lobes of MLAN3.
This system represents one of the first samples of giant double-lobed radio galaxies associated with enormous Ly$\alpha$ nebulae at high-$z$.
Please refer to our upcoming work (Li et al. in prep) for a detailed analysis of this target.

\begin{deluxetable}{cc}
\tablenum{3}
\tablecaption{Measured fluxes of stacked nebulae\label{tab:stack}}
\tablewidth{0pt}
\tablehead{
\colhead{Filter Band}  & \colhead{Flux Density $f_\nu$ [$\rm \mu Jy$]}
}
\startdata
HSC g & $1.22\pm0.26$\tablenotemark{$\dagger$}\\
WISE W1 3.4$~\mu$m & $29\pm8$\\
WISE W2 4.6$~\mu$m & $37\pm11$\\
WISE W3 12$~\mu$m & $286\pm32$\\
WISE W4 22$~\mu$m & $690\pm250$ \\
VLA S 2-4GHz & $19\pm6$\\
\enddata
\tablenotetext{\dagger}{AB magnitude $\mathrm{m_g}=23.68\pm0.23$.}
\end{deluxetable}

\begin{figure*}[t]
    \centering
    \includegraphics[width=\linewidth]{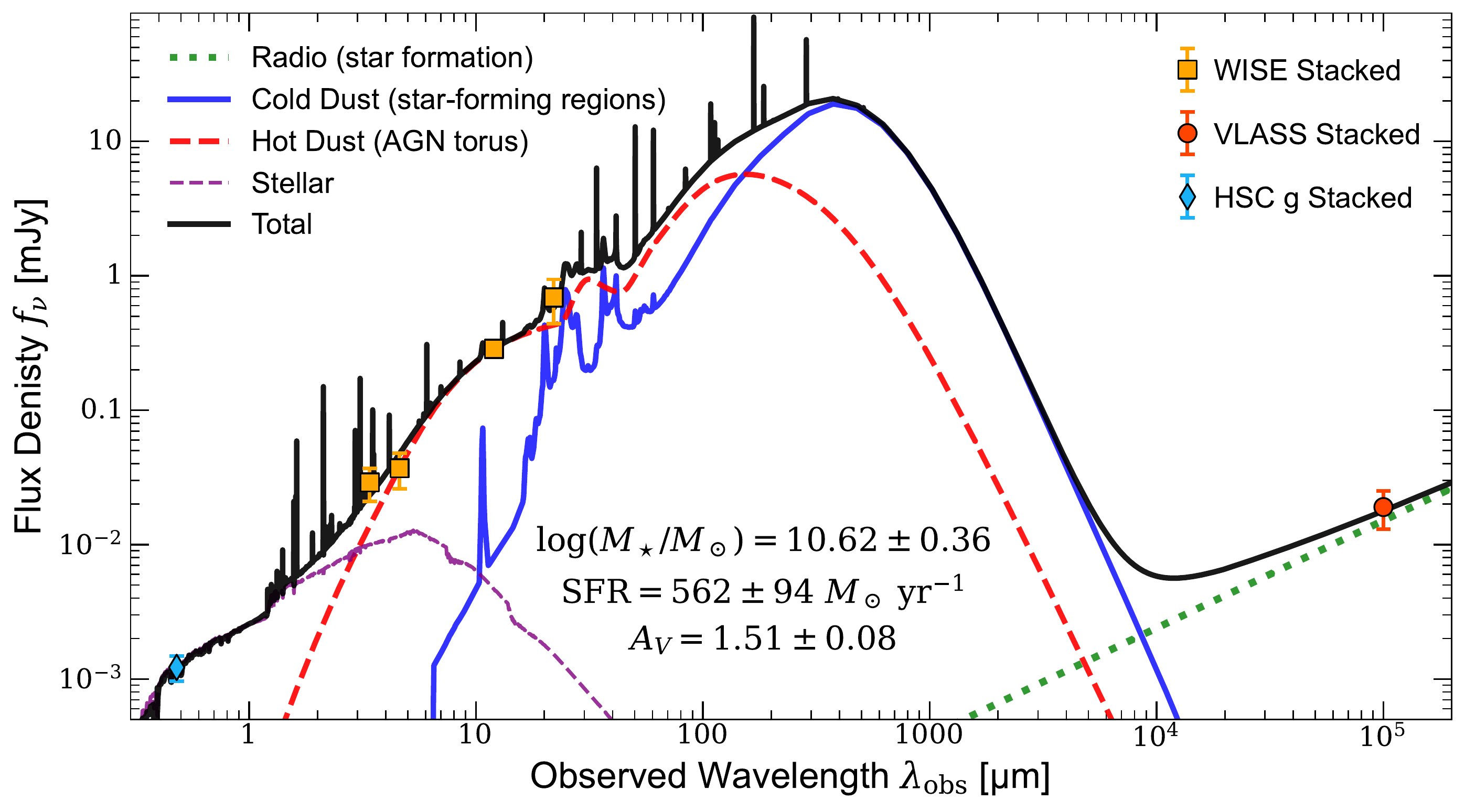}
    \caption{Spectral energy distribution (SED) for the average source powering Ly$\alpha$ nebulae by stacking all 117 targets in this work. The observed data are from stacking of our Subaru/HSC g-band imaging (blue diamond), WISE imaging (orange squares), and VLASS S-band imaging (red circle). The black line denotes the best fit of \texttt{CIGALE}, together with the individual components: hot dust emission (red dashed line), cold dust emission (blue line), stellar emission (purple dashed line), and the radio emission from star formation (green dotted line). The best-fit galaxy stellar mass, SFR, and dust attenuation are annotated in the figure.}
    \label{fig:SED}
\end{figure*}

\subsection{Existence of active galactic nuclei}
\label{sec:agn}

Despite the fact that detailed Ly$\alpha$ powering mechanisms are still debated, there is a consensus that the role of AGN is not negligible, especially for the brightest Ly$\alpha$ nebulae \citep{Geach2009}.
However, except for the ultra-luminous broad-line QSO, it is challenging to identify AGNs unambiguously, especially for our high redshift sample. 
In this section, we discuss the emergence of AGN in our nebula sample.

A few approaches can reveal the presence of an AGN.
The most robust evidence is X-ray detection.
We checked for X-ray counterparts to our nebula sample using the XMM-Newton Serendipitous Source Catalogue \citep{Webb2020} and Chandra Source Catalog \citep{Evans2010}.
Because our survey fields were newly targeted, no deep X-ray observation covered them.
Only partial overlap exists for fields J0222 and J0210, but no X-ray counterparts were detected for any of our nebulae in either catalog.
Radio emissions can also act as evidence of AGN activity, despite only a small fraction of AGN is radio-loud, probably due to a short duty cycle of the radio-loud phase \citep[\eg,][]{Tadhunter2016}.
As the section~\ref{sec:radio} presents, two nebulae (MLAN3 and MLAN44) have mJy-level radio counterparts.
Also note that MLAN44 has extremely bright mid-IR fluxes, indicating a dust-obscured AGN's existence.

Emission line properties from spectroscopy can help identify the emergence of AGN.
We can estimate an unbiased AGN fraction in our sample based on our spectroscopic follow-up, which requires no prior knowledge of the AGN activity.
Out of seven confirmed nebulae (Figure~\ref{fig:spec}), five show AGN features (broad \ion{C}{4} lines for MLAN14, MLAN64, and MLAN84; highly ionized \ion{N}{5} lines for MLAN1; radio galaxy for MLAN3).
This implies an AGN fraction of $5/7 \simeq 71\%$.
Moreover, our sample also includes QSOs targeted by spectroscopic surveys.
Among the 117 nebulae in our sample, eight (MLAN7, MLAN8, MLAN14, MLAN16, MLAN18,  MLAN27, MLAN28, and MLAN31) have been surveyed and confirmed as QSOs by SDSS \citep{Lyke2020}.


\subsection{Understanding the Observed Diversity: Manifestations of Massive Galaxy Evolution}
\label{sec:unify}

The population of Ly$\alpha$ nebulae displays a remarkable diversity in their observed multi-wavelength properties.
By synthesizing the observations across the multiple wavebands, we unravel the intricate natures of nebulae and associated galaxies.
Despite this diversity, there is supposed to be an overarching picture of galaxy evolution driving the emergence of extended Ly$\alpha$ nebulae.
In this section, we aim to understand the observed diversity of Ly$\alpha$ nebulae under the framework of massive galaxy formation and evolution.

In order to gain an average view of host galaxies of Ly$\alpha$ nebulae, we stack imaging data in the rest-frame UV (Subaru/HSC g-band), IR (WISE W1,W2,W3,W4), and radio (VLASS S-band) wavelengths.
We stack the cutout images of all sources by clipping the 3$\sigma$ outliers and taking the mean on the pixel scale.
These bands can trace key properties including dust attenuation, star formation rate, and AGN activity.
The stacked imaging of all these bands show $>2\sigma$ signals.
The measured fluxes are listed in Table~\ref{tab:stack}.
Note that we test the stacking by removing QSOs, radio galaxies, or dust-obscured galaxies (MLAN44), which give the same results as the stacking of the full sample to within 1$\sigma$..
Therefore, aiming to get the average property of Ly$\alpha$ nebulae, we include all 117 sample sources in the stacking process.

We run spectral energy distribution (SED) fitting for the stacked source using \texttt{CIGALE}, which is a Python-based code to efficiently model the X-ray to radio spectrum of galaxies with AGN and estimate their physical properties such as star formation rate, attenuation, and stellar mass \citep{Boquien2019, Yang2022}.
For \texttt{CIGALE} modeling, we assume a delayed-$\tau$ star-formation history, where an optional late starburst is allowed in the last 20 Myr.
We use \citet{Bruzual2003} stellar population synthesis models and a broad ionization parameter range of -1.0 to -3.0.
We adopt a modified \citet{Calzetti2000} attenuation curve, which allows the variation of the power-law slope by $\pm0.3$.
We use the dust emission templates of \citet{Dale2014} and AGN models from \citet{Stalevski2012, Stalevski2016}.
Besides, galaxy synchrotron and AGN radio emissions are included.

The best-fit model and physical properties are presented in Fig.~\ref{fig:SED}.
The SED reveals a massive galaxy ($\log M_\star/M_\odot = 10.62\pm0.36$) with copious dust ($A_V=1.51\pm0.08$), violent star formation (SFR=$562\pm94~M_\odot~\mathrm{yr}^{-1}$), and obscured AGN.
This clearly shows the dusty nature of the host galaxy associated with the Ly$\alpha$ nebula.
The dusty nature of galaxies associated with Ly$\alpha$ nebulae implies that copious Ly$\alpha$ photons and UV photons can escape from the dusty galaxies to the CGM scale.
While the obscuration makes Ly$\alpha$ and UV escape difficult, some can surely escape from dusty galaxies through mechanisms such as geometric (anisotropic escape), kinematic (wind dispersal), and radiative transfer effects.
Otherwise, luminous \textit{in-situ} Ly$\alpha$ emissions are needed in the CGM.
Further studies are expected to resolve this puzzle of luminous Ly$\alpha$ nebulae around dusty galaxies.

Synthesizing multi-wavelength observations, we propose that Ly$\alpha$ nebulae can represent a general manifestation during the key phase of massive galaxy assembly.
The diverse population of Ly$\alpha$ nebulae, therefore, provides a window into the intricate life cycles of massive galaxies emerging from dust during the peak epoch of galaxy assembly \cite[\eg,][]{Hopkins2008}.
Early in the starburst stage, galaxies undergo intense dust-obscured starburst activity, and energetic AGN turn on at the galaxy center driving large-scale gas flow and powering extended Ly$\alpha$ nebulae.
This dusty phase dominates the time scale and represents an average powering source as shown by our SED fitting of stacked Ly$\alpha$ nebulae (Fig.~\ref{fig:SED}).
As the galaxies evolve, they will experience a very short-time ``blowout'' phase, possibly observed as radio-loud galaxies.
Further, with the copious dust and gas exhausted and expelled, the QSO will be bare and then decay.
The small fraction of radio galaxy ($\lesssim 2\%$) and QSO ($\lesssim 7\%$) can be well explained by the short time scale at these stages.
The majority ($\sim80\%$) of enormous Ly$\alpha$ nebulae is Type II, whose galaxies are hidden in dust and UV-faint, with violent starburst ongoing and possible obscured AGN shining.
Further, the dominant dusty nature of Ly$\alpha$ nebula shown by our stacked results proves that a long dusty lifetime occurs when the massive galaxies undergo violent starburst and black hole assembly.
In the future, IR and (sub)mm observations (\eg, JWST and ALMA) for the discovered Ly$\alpha$ nebulae and Ly$\alpha$ observations for dusty starburst galaxies and obscured AGNs at high-$z$ would be critical for disentangling the histories of individual systems and piecing together an overarching picture of massive galaxy formation and evolution at the cosmic noon.

\section{Summary}
\label{sec:summary}
We have carried out wide-field, deep imaging observations of eight fields selected by the MAMMOTH approach at $z=2.18$ and $z=2.29$ using Subaru/HSC.
We select 117 Ly$\alpha$ nebulae in an unbiased way and compile multi-wavelength data from UV to radio to achieve a panchromatic perspective on the properties of Ly$\alpha$ nebulae.
An observed diversity of nebula population has been revealed.
The main conclusions are summarized as follows:
\begin{enumerate}
    \item Searching over 12 deg$^2$, we discover 117 Ly$\alpha$ nebulae with luminous ($L_\mathrm{Ly\alpha}=8 - 40\times10^{43}~\rm erg~s^{-1}$) and extended (40 - 400 kpc) Ly$\alpha$ emissions.
    Among the sample, 28 nebulae have Ly$\alpha$ sizes larger than 100 kpc and are labeled as enormous Ly$\alpha$ nebulae (ELANe).
    We confirm the redshifts of seven nebulae spectroscopically, with an AGN fraction ($\gtrsim71\%$) indicated by broad emission lines and high-ionization emission lines.
    \item An intergalactic nebula, dubbed the Ivory Nebula, is highlighted with extended filamentary and well-resolved clumpy Ly$\alpha$ emission between two galaxies.
    The Ivory Nebula has a Ly$\alpha$ luminosity of $L_\mathrm{Ly\alpha}=2.5\times10^{44}~\rm erg~s^{-1}$ and an end-to-end angular extent of $44.4\arcsec \simeq 365 ~\rm kpc$.
    The extreme properties make it one of the largest nebulae discovered in the Universe and provide an opportunity to directly study gas flow in galaxy formation.
    \item We present an observed diversity of Ly$\alpha$ nebula multi-wavelength properties.
    There is a bimodal distribution for UV magnitude of powering galaxies.
    UV-faint sources ($M_\mathrm{UV}>-22$) dominate the population of enormous Ly$\alpha$ nebulae with size $>100~\rm kpc$.
    This intriguing class of Ly$\alpha$ nebulae is categorized as Type II ELAN.
    Type II ELANe constitute the majority of enormous nebulae in the universe but remain largely hidden in all-sky surveys.
    Dusty starburst and obscured AGN activity can help to resolve the puzzle.
    \item A small fraction of Ly$\alpha$ nebulae have very luminous IR counterparts ($\sim1\%$ $\rm W4_{24\mu m}>5~mJy$) and radio galaxy counterparts ($\sim2\%$).
    About $7\%$ Ly$\alpha$ nebulae are associated with QSOs.
    Radio galaxy nebulae could trace a nebula-rich large-scale environment.
    Out of 28 ELANe in our sample, 18 ($\sim64\%$) are located in the J0240 and J1349 fields, where two radio nebulae reside.
    \item The SED of stacked Ly$\alpha$ nebulae reveal a massive ($\log M_\star/M_\odot = 10.62\pm0.36$), dusty ($A_V=1.51\pm0.08$), star-forming ($\rm SFR=562\pm94~M_\odot~yr^{-1}$) galaxy with an obscured AGN activity.
\end{enumerate}

We propose a model where the diverse Ly$\alpha$ nebula categories capture massive galaxies at different evolutionary stages as they are violently assembling.
A relatively long dusty stage has been suggested by the majority of Type II ELANe and dusty nature of stacked SED.
In this stage, violent star formation and obscured AGN can play an role in powering the circumgalactic Ly$\alpha$ nebulae.
In conclusion, we collect a large sample of Ly$\alpha$ nebulae and underscore the pivotal role of Ly$\alpha$ nebulae in understanding the intricate processes governing galaxy formation and evolution at high-$z$.
In the future, more infrared observations, such as those from the James Webb Space Telescope (JWST) and the Atacama Large Millimeter/submillimeter Array (ALMA), will be vitally needed for studying the discovered Ly$\alpha$ nebulae.
Additionally, observing Ly$\alpha$ emissions of more dusty starburst galaxies and obscured AGNs at high-$z$ in an unbiased way will be essential for unraveling the physical processes of these systems.
Together, these efforts will paint an overarching picture of the formation and evolution of massive galaxies during the cosmic noon.

\section*{Acknowledgments}

M.Li, Z.Cai, Y.Wu, S.Zhang, and B.Wang acknowledge support from the National Key R\&D Program of China (grant no.~2023YFA1605600), the National Science Foundation of China (grant no.~12073014), the science research grants from the China Manned Space Project with No.~CMS-CSST-2021-A05, and Tsinghua University Initiative Scientific Research Program (No.~20223080023). SC gratefully acknowledges support from the European Research Council (ERC) under the European Union’s Horizon 2020 Research and Innovation programme grant agreement No 864361.”


This research is based in part on data collected by Hyper Suprime-Cam at the Subaru Telescope, which is operated by the National Astronomical Observatory of Japan. We are honored and grateful for the opportunity of observing the Universe from Maunakea, which has the cultural, historical, and natural significance in Hawaii.
The authors wish to recognize and acknowledge the very significant cultural role and reverence that the summit of Maunakea has always had within the indigenous Hawaiian community.  We are most fortunate to have the opportunity to conduct observations from this mountain.
The Hyper Suprime-Cam (HSC) collaboration includes the astronomical communities of Japan, Taiwan, and Princeton University. The HSC instrumentation and software were developed by the National Astronomical Observatory of Japan (NAOJ), the Kavli Institute for the Physics and Mathematics of the Universe (Kavli IPMU), the University of Tokyo, the High Energy Accelerator Research Organization (KEK), the Academia Sinica Institute for Astronomy and Astrophysics in Taiwan (ASIAA), and Princeton University. Funding was contributed by the FIRST program from Japanese Cabinet Office, the Ministry of Education, Culture, Sports, Science and Technology (MEXT), the Japan Society for the Promotion of Science (JSPS), Japan Science and Technology Agency (JST), the Toray Science Foundation, NAOJ, Kavli IPMU, KEK, ASIAA, and Princeton University. 

This paper makes use of software developed for the Large Synoptic Survey Telescope. We thank the LSST Project for making their code available as free software at \url{http://dm.lsst.org}

This paper makes use of data from PS1. The Pan-STARRS1 Surveys (PS1) have been made possible through contributions of the Institute for Astronomy, the University of Hawaii, the Pan-STARRS Project Office, the Max-Planck Society and its participating institutes, the Max Planck Institute for Astronomy, Heidelberg and the Max Planck Institute for Extraterrestrial Physics, Garching, The Johns Hopkins University, Durham University, the University of Edinburgh, Queen’s University Belfast, the Harvard-Smithsonian Center for Astrophysics, the Las Cumbres Observatory Global Telescope Network Incorporated, the National Central University of Taiwan, the Space Telescope Science Institute, the National Aeronautics and Space Administration under Grant No. NNX08AR22G issued through the Planetary Science Division of the NASA Science Mission Directorate, the National Science Foundation under Grant No. AST-1238877, the University of Maryland, and Eotvos Lorand University (ELTE) and the Los Alamos National Laboratory.

This publication makes use of data products from the Wide-field Infrared Survey Explorer, which is a joint project of the University of California, Los Angeles, and the Jet Propulsion Laboratory/California Institute of Technology, funded by the National Aeronautics and Space Administration.

This paper includes data gathered with the 6.5-meter Magellan Telescope located at Las Campanas Observatory, Chile.


\appendix
\section{Approach to estimate the line flux from two-band observations}

\label{sec:estimate_line}
In this appendix, we delineate the methodology for calculating the flux of emission lines and the continuum from multi-band imaging data.

We first characterize the basic notation and definition.
The intrinsic spectral energy distribution of an astrophysical source as the radiative flux density per unit frequency or wavelength interval is described by $f_\nu$ or $f_\lambda$, expressed in $\nu$-units $(\rm erg~cm^{-2}~s^{-1}~Hz^{-1})$ and $\lambda$-units $(\rm erg~cm^{-2}~s^{-1}~\AA^{-1})$, respectively.
These two representations are related through:
$$
f_\nu\mathrm{~d} \nu=f_\lambda \mathrm{~d} \lambda \quad; \quad c = \lambda \nu\quad ; \quad |\frac{\mathrm{d} \nu}{\mathrm{~d} \lambda}|=\frac{\mathrm{c}}{\lambda^2},
$$
where $c$ is the speed of light in vacuum.

Photometric measurements typically utilize photometric filters that probe the flux density over defined wavelength intervals or passbands \citep[\eg,][]{Bessell1990, Fukugita1996}.
Photometric filters are defined by their transmission curves $T_\mathrm{x}(\lambda)$, which describe their response as a function of wavelength.
For convenience, we use $\mathrm{T}_\mathrm{x}(\lambda)$ to present the total system response including atmospheric transmission, the throughput of the telescope, the instrument, the quantum efficiency of the detector, the filter response function, and any other response.
All photons received within a given pass-band during the measuring process get integrated, hence the wavelength distribution of $f_\lambda$ are lost.
For this reason, the flux of a source measured in a given filter ``x'' is effectively defined as the average flux density in the pass-band weighted by the filter transmission curve.
For photon-counting detectors like CCDs, The number of photoelectrons per second detected by filter ``x'' from a source with an intrinsic spectral energy distribution $f_\lambda(\lambda)$ is
\begin{equation}
N_p =\int f_\lambda(\lambda) T_{\rm x}(\lambda) / h \nu ~{\rm d} \lambda 
=\frac{1}{h c} \int \lambda f_\lambda(\lambda) T_{\rm x}(\lambda) ~\rm d \lambda,
\label{eq:photon}
\end{equation}
where $T_{\rm x}(\lambda)$ is the total system response.
If $f_{\lambda}(\lambda)$ and $T_{\rm x}(\lambda)$ are both continuous and $T_{\rm x}(\lambda)$ is nonnegative over the wavelength interval, then from equation (\ref{eq:photon}) and the Mean Value Theorem for Integrals there exists a $\lambda_{\rm iso}$ such that 
$
f_\lambda\left(\lambda_{\text {iso}}\right) \int \lambda T_{\rm x}(\lambda) \mathrm{~d} \lambda=\int \lambda f_\lambda(\lambda) T_{\rm x}(\lambda) \mathrm{~d} \lambda.
$
Thus we obtain
\begin{equation}
\label{eq:flambda}
\left\langle f_\lambda\right\rangle_{\rm x}=f_\lambda\left(\lambda_{\text {iso}}\right)=\frac{\int  f_\lambda T_{\rm x} \lambda\mathrm{~d} \lambda}{\int  T_{\rm x} \lambda\mathrm{~d} \lambda},
\end{equation}
where $\lambda_{\rm iso}$ denotes the ``isophotal wavelength'' and $\left\langle f_\lambda\right\rangle_{\rm x}$ denotes the \textbf{mean flux density} measured by filter band ``x''.

We can describe equation (\ref{eq:photon}) in terms of frequency.
Note that ${\rm d}\lambda/\lambda = {\rm d}\nu/\nu$ and $f_{\lambda} = f_{\nu}\frac{c}{\lambda^2}$.
Thus we have
\begin{equation}
N_p=\frac{1}{h c} \int \lambda f_\lambda(\lambda) T_{\rm x}(\lambda) {\rm ~d} \lambda=\frac{1}{h} \int f_\nu(\nu) T_{\rm x}(\nu) {\rm ~d} \nu / \nu.
\end{equation}
Similar to the derivation given by equation (\ref{eq:flambda}), we can also define the mean flux density in $\nu$-units:
\begin{equation}
\left\langle f_\nu\right\rangle_\mathrm{x}
=\frac{\int f_\nu T_\mathrm{x} \frac{\mathrm{d} \nu}{\nu}}{\int T_\mathrm{x} ~\frac{\mathrm{d} \nu}{\nu}}
=\frac{\int f_\lambda T_\mathrm{x} \lambda \mathrm{d} \lambda}{c\int T_\mathrm{x} ~\frac{\mathrm{d} \lambda}{\lambda}}.
\end{equation}
For consistency of conversion, we need a wavelength $\lambda_\mathrm{pivot,x}$ to make
\begin{equation}
\label{eq:pivot_wavelength_conversion}
\langle f_{\nu} \rangle_\mathrm{x} = \frac{\lambda_{\rm pivot, x}^2}{c} \langle f_{\lambda} \rangle_\mathrm{x}.
\end{equation}
We obtain
\begin{equation}
    \lambda_\mathrm{pivot,x} = \sqrt{\frac{\int T_\mathrm{x}\lambda\mathrm{~d \lambda}}{\int T_\mathrm{x}\frac{\mathrm{d}\lambda}{\lambda}}}.
\end{equation}
The $\lambda_{\rm pivot, x}$ is known as the \textbf{pivot wavelength} of filter band ``x'' \citep[\eg,][]{Koornneef1986,Tokunaga2005}.
The pivot wavelength provides an exact relation between $\langle f_{\nu} \rangle_\mathrm{x}$ and $\langle f_{\lambda} \rangle_\mathrm{x}$ by equation (\ref{eq:pivot_wavelength_conversion}).
For convenience, we denote $\langle \lambda^{\alpha} \rangle_\mathrm{x} = \int \lambda^{\alpha} T_{\rm x}{\rm ~d} \lambda/\int T_{\rm x}{\rm ~d} \lambda$.
Hence the pivot wavelength can be written as $\lambda_\mathrm{pivot,x} = \sqrt{{\langle \lambda\rangle_\mathrm{x}}/{\langle \lambda^{-1} \rangle_\mathrm{x}}}$.

We assume the spectral energy distribution of a line-emitting sources is a power-law continuum plus a delta function emission line:
\begin{equation}
    f_\lambda = f_{\rm \lambda, cont} + f_{\rm \lambda, line} = f_0 (\lambda/\lambda_0)^\beta + F_{\rm line}\delta(\lambda-\lambda_{\rm line}),
\end{equation}
where $\beta$ is the slope of the power-law continuum, $F_{\rm line}$ is the line flux, and $\lambda_{\rm line}$ is the line wavelength.

Using the filter "x" to "measure" average flux density for this mock spectrum, we get
\begin{equation}
\langle f_{\lambda} \rangle_\mathrm{x}
= \frac{\int f_\lambda T_{\rm x} \lambda \mathrm{~d} \lambda}{\int T_{\rm x} \lambda \mathrm{~d} \lambda}
= \frac{\int f_0 (\lambda/\lambda_0)^\beta T_{\rm x} \lambda \mathrm{~d} \lambda}{\int T_{\rm x} \lambda \mathrm{~d} \lambda} + \frac{F_{\rm line} T_{\rm x}(\lambda_{\rm line}) \lambda_{\rm line}}{\int T_{\rm x} \lambda \mathrm{~d} \lambda}
= \frac{f_0}{\lambda_0^\beta}\frac{\langle\lambda^{\beta+1}\rangle_{\rm x}}{\langle\lambda\rangle_{\rm x}} + \frac{F_{\rm line}}{\Delta_{\rm x}(\lambda_{\rm line})} \frac{\lambda_{\rm line}}{\langle\lambda\rangle_{\rm x}},
\end{equation}
where $\Delta_{\rm x}(\lambda) \equiv \frac{\int T_{\rm x}{~\rm d}\lambda}{T_{\rm x}(\lambda)}$ is the effective band width of filter ``x'' at wavelength $\lambda$.
Note that this effective band width depends on the wavelength, which equals to the horizontal size of a rectangle with a height equal to
the transmission at a given wavelength $\lambda$ and with the same area that the one covered by the filter transmission curve.

For the two-band scenario, we use a narrowband (NB) and broadband (BB) to capture the line-emitting sources.
Therefore, we get linear equations:
\begin{equation}
\left\{
\begin{aligned}
\langle f_{\lambda} \rangle_\mathrm{NB} &= \frac{f_0}{\lambda_0^\beta}\frac{\langle\lambda^{\beta+1}\rangle_{\rm NB}}{\langle\lambda\rangle_{\rm NB}} + \frac{F_{\rm line}}{\Delta_{\rm NB}(\lambda_{\rm line})} \frac{\lambda_{\rm line}}{\langle\lambda\rangle_{\rm NB}} \\
\langle f_{\lambda} \rangle_\mathrm{BB} &= \frac{f_0}{\lambda_0^\beta}\frac{\langle\lambda^{\beta+1}\rangle_{\rm BB}}{\langle\lambda\rangle_{\rm BB}} + \frac{F_{\rm line}}{\Delta_{\rm BB}(\lambda_{\rm line})} \frac{\lambda_{\rm line}}{\langle\lambda\rangle_{\rm BB}} \\
\end{aligned}
\right..
\end{equation}
The solution is
\begin{equation}
\left\{
\begin{aligned}
f_{\rm \lambda, cont}(\lambda) &= f_0(\lambda/\lambda_0)^\beta = \frac{\langle f_{\lambda} \rangle_\mathrm{NB}\Delta_{\rm NB}(\lambda_{\rm line})\langle \lambda\rangle_\mathrm{NB} 
\lambda_{\rm line}^\beta- \langle f_{\lambda} \rangle_\mathrm{BB}\Delta_{\rm BB}(\lambda_{\rm line})\langle \lambda\rangle_\mathrm{BB}\lambda_{\rm line}^\beta}{\Delta_{\rm NB}(\lambda_{\rm line})\langle \lambda^{\beta+1}\rangle_\mathrm{NB}
 - \Delta_{\rm BB}(\lambda_{\rm line})\langle \lambda^{\beta+1}\rangle_\mathrm{BB}
}(\lambda/\lambda_{\rm line})^\beta \\
F_{\rm line} &= \left({\langle f_{\lambda} \rangle_\mathrm{NB}\frac{\langle \lambda\rangle_\mathrm{NB}}{\langle \lambda^{\beta+1}\rangle_\mathrm{NB}}-\langle f_{\lambda} \rangle_\mathrm{BB}\frac{\langle \lambda\rangle_\mathrm{BB}}{\langle \lambda^{\beta+1}\rangle_\mathrm{BB}}}\right)\Big/\left({\frac{\lambda_{\rm line}}{\Delta_{\rm NB}(\lambda_{\rm line})\langle \lambda^{\beta+1}\rangle_\mathrm{NB}} - \frac{\lambda_{\rm line}}{\Delta_{\rm BB}(\lambda_{\rm line})\langle \lambda^{\beta+1}\rangle_\mathrm{BB}}}\right)\\
\end{aligned}
\right..
\end{equation}

Because AB magnitude is preferred in galaxy photometry, which is defined as $m_{\rm AB} = -2.5\log(f_\nu [\mathrm{erg~s^{-1}~{cm^{-2}}~Hz^{-1}}]) - 48.6$, we present the results in $f_\nu$ converting by pivot wavelength:
\begin{equation}
\left\{
\begin{aligned}
f_{\rm \lambda, cont}(\lambda) &= f_0(\lambda/\lambda_0)^\beta = \frac{\langle f_{\nu} \rangle_\mathrm{NB}\Delta_{\rm NB}(\lambda_{\rm line})\langle \lambda^{-1}\rangle_\mathrm{NB} 
\lambda_{\rm line}^\beta- \langle f_{\nu} \rangle_\mathrm{BB}\Delta_{\rm BB}(\lambda_{\rm line})\langle \lambda^{-1}\rangle_\mathrm{BB}\lambda_{\rm line}^\beta}{\Delta_{\rm NB}(\lambda_{\rm line})\langle \lambda^{\beta+1}\rangle_\mathrm{NB}
 - \Delta_{\rm BB}(\lambda_{\rm line})\langle \lambda^{\beta+1}\rangle_\mathrm{BB}
}c(\lambda/\lambda_{\rm line})^\beta \\
F_{\rm line} &= c\left({\langle f_{\nu} \rangle_\mathrm{NB}\frac{\langle \lambda^{-1}\rangle_\mathrm{NB}}{\langle \lambda^{\beta+1}\rangle_\mathrm{NB}}-\langle f_{\nu} \rangle_\mathrm{BB}\frac{\langle \lambda^{-1}\rangle_\mathrm{BB}}{\langle \lambda^{\beta+1}\rangle_\mathrm{BB}}}\right)\Big/\left({\frac{\lambda_{\rm line}}{\Delta_{\rm NB}(\lambda_{\rm line})\langle \lambda^{\beta+1}\rangle_\mathrm{NB}} - \frac{\lambda_{\rm line}}{\Delta_{\rm BB}(\lambda_{\rm line})\langle \lambda^{\beta+1}\rangle_\mathrm{BB}}}\right)\\
\end{aligned}
\right..
\end{equation}

To simplify the equations, we define the coefficients as
\begin{equation}
\begin{aligned}
\epsilon_{\rm cont, NB} &= \frac{\Delta_{\rm NB}(\lambda_{\rm line})\langle \lambda^{-1}\rangle_\mathrm{NB} 
\lambda_{\rm line}^{\beta+2}}{\Delta_{\rm BB}(\lambda_{\rm line})\langle \lambda^{\beta+1}\rangle_\mathrm{BB}
 - \Delta_{\rm NB}(\lambda_{\rm line})\langle \lambda^{\beta+1}\rangle_\mathrm{NB}},\\
 \epsilon_{\rm cont, BB} &= \frac{\Delta_{\rm BB}(\lambda_{\rm line})\langle \lambda^{-1}\rangle_\mathrm{BB} 
\lambda_{\rm line}^{\beta+2}}{\Delta_{\rm BB}(\lambda_{\rm line})\langle \lambda^{\beta+1}\rangle_\mathrm{BB}
 - \Delta_{\rm NB}(\lambda_{\rm line})\langle \lambda^{\beta+1}\rangle_\mathrm{NB}},\\
 \epsilon_{\rm line, NB} &= \frac{\lambda_{\rm line}\langle \lambda^{-1}\rangle_\mathrm{NB}}{\langle \lambda^{\beta+1}\rangle_\mathrm{NB}}\Big/\left({ \frac{1}{\Delta_{\rm NB}(\lambda_{\rm line})\langle \lambda^{\beta+1}\rangle_\mathrm{NB}}-\frac{1}{\Delta_{\rm BB}(\lambda_{\rm line})\langle \lambda^{\beta+1}\rangle_\mathrm{BB}} }\right),\\
  \epsilon_{\rm line, BB} &= \frac{\lambda_{\rm line}\langle \lambda^{-1}\rangle_\mathrm{BB}}{\langle \lambda^{\beta+1}\rangle_\mathrm{BB}}\Big/\left({\frac{1}{\Delta_{\rm NB}(\lambda_{\rm line})\langle \lambda^{\beta+1}\rangle_\mathrm{NB}}-\frac{1}{\Delta_{\rm BB}(\lambda_{\rm line})\langle \lambda^{\beta+1}\rangle_\mathrm{BB}} }\right).\\
\end{aligned}
\end{equation}

Then the solution will be
\begin{equation}
\left\{
\begin{aligned}
f_{\rm \lambda, cont}(\lambda) &= \left(\epsilon_{\rm cont, BB} \langle f_{\nu} \rangle_\mathrm{BB} - \epsilon_{\rm cont, NB} \langle f_{\nu} \rangle_\mathrm{NB} \right) \frac{c}{\lambda_{\rm line}^2} (\lambda/\lambda_{\rm line})^\beta\\
F_{\rm line} &= \left(\epsilon_{\rm line, NB} \langle f_{\nu} \rangle_\mathrm{NB} - \epsilon_{\rm line, BB} \langle f_{\nu} \rangle_\mathrm{BB} \right) \frac{c}{\lambda_{\rm line}^2}
\end{aligned}
\right..
\end{equation}



In this work, we assume a constant continuum $f_\nu$, which refers to $\beta = -2$.
Therefore, the coefficients can be simplified to be:
\begin{equation}
\begin{aligned}
\epsilon_{\rm cont, NB} &= \frac{\Delta_{\rm NB}(\lambda_{\rm line})\langle \lambda^{-1}\rangle_\mathrm{NB}}{\Delta_{\rm BB}(\lambda_{\rm line})\langle \lambda^{-1}\rangle_\mathrm{BB}
 - \Delta_{\rm NB}(\lambda_{\rm line})\langle \lambda^{-1}\rangle_\mathrm{NB}},\\
 \epsilon_{\rm cont, BB} &= 1+\epsilon_{\rm cont, NB}\\
 \Delta_{\rm NB; BB|\lambda_{line}} \equiv \epsilon_{\rm line, NB} = \epsilon_{\rm line, BB} &= \lambda_{\rm line}\left({\frac{1}{\Delta_{\rm NB}(\lambda_{\rm line})\langle \lambda^{-1}\rangle_\mathrm{NB}}-\frac{1}{\Delta_{\rm BB}(\lambda_{\rm line})\langle \lambda^{-1}\rangle_\mathrm{BB}} }\right)^{-1},\\
\end{aligned}
\end{equation}
where $\Delta_{\rm NB; BB|\lambda_{line}}$ is the calibrated effective bandwidth.
The solutions in the assumption can be
\begin{equation}
\left\{
\begin{aligned}
f_{\rm \nu, cont} &= \epsilon_{\rm cont, BB} \langle f_{\nu} \rangle_\mathrm{BB} - \epsilon_{\rm cont, NB} \langle f_{\nu} \rangle_\mathrm{NB}\\
F_{\rm line} &= \left(\langle f_{\nu} \rangle_\mathrm{NB} - \langle f_{\nu} \rangle_\mathrm{BB} \right) \frac{c\Delta_{\rm NB; BB|\lambda_{line}}}{\lambda_{\rm line}^2}
\end{aligned}
\right.
\end{equation}


We use the pivot wavelength of the NB as the line wavelength.
For the NB400 and g-band, the coefficients are:
\begin{equation}
\begin{aligned}
\epsilon_{\rm cont, NB400} &= 0.027,\\
 \epsilon_{\rm cont, g} &= 1.027,\\
 \epsilon_{\rm line, NB400} = \epsilon_{\rm line, g} &= 101.0~\AA.\\
\end{aligned}
\end{equation}
For the NB387 and g-band, the coefficients are:
\begin{equation}
\begin{aligned}
\epsilon_{\rm cont, NB387} &= 0,\\
 \epsilon_{\rm cont, g} &= 1,\\
 \epsilon_{\rm line, NB387} = \epsilon_{\rm line, g} &= 55.9~\AA.\\
\end{aligned}
\end{equation}


\bibliography{mingyu, tmp}{}
\bibliographystyle{aasjournal}

\suppressAffiliationsfalse
\allauthors 

\end{document}